\documentclass[final,3p,times,twocolumn]{elsarticle}

\usepackage{blindtext, graphicx, amsmath, algorithm, algpseudocode, pifont, algcompatible, comment, layout, amsthm, amssymb}
\usepackage{enumitem}   
\usepackage{eso-pic}
\usepackage{braket}
\usepackage{booktabs,tabularx}
\usepackage{float}
\usepackage{makecell}
\usepackage{subcaption} 

\usepackage[utf8]{inputenc}
\usepackage[english]{babel}
\usepackage[hidelinks]{hyperref} 
\hypersetup{ colorlinks=true, linkcolor=black, filecolor=black, urlcolor=cyan, }

\usepackage{caption}
\makeatletter

\journal{ICT Express}

\begin{document}

\begin{frontmatter}

\title{A Survey of OAM-Encoded High-Dimensional Quantum Key Distribution: Foundations, Experiments, and Recent Trends}

\author{Huan Zhang}
\ead{cola1999@hanyang.ac.kr}

\author{Zhenyu Cao}
\ead{zycao@hanyang.ac.kr}

\author{Yu Sun}
\ead{sunyu0817@hanyang.ac.kr}

\author{Hu Jin\corref{cor1}}
\ead{hjin@hanyang.ac.kr}

\address{Department of Electrical and Electronic Engineering, Hanyang University, Ansan, South Korea}

\cortext[cor1]{Corresponding author}

\begin{abstract}
High-dimensional quantum key distribution (HD-QKD) enhances information efficiency and noise tolerance by encoding data in large Hilbert spaces. The orbital angular momentum (OAM) of light provides a scalable basis for such encoding and supports high-dimensional photonic communication. Practical OAM-based implementations remain constrained by challenges in state generation, transmission, and detection. This survey offers a consolidated overview of OAM-encoded HD-QKD, outlining fundamental principles, representative experiments, and system-level limitations. Recent progress in hybrid encodings, mode sorting, adaptive optics, and TF, CV, MDI, and DI frameworks is summarized with emphasis on practical feasibility.
\end{abstract}


\end{frontmatter}

\section{Introduction}\label{sec_intro}
With the rapid progress of quantum computing \cite{shor1999polynomial}, which threatens the security foundations of classical public-key cryptosystems, the global communications infrastructure increasingly recognizes quantum key distribution (QKD) as a core technology for ensuring long-term confidentiality. Reflecting this shift, by 2025 the ITU-T has updated the baseline specification for QKD control and management \cite{ITU-T-Y3804-2025} and, in 2024, finalized interworking standards \cite{ITU-T-Y3820-2024} along with a dedicated security recommendation \cite{ITU-T-X1717-2024}. QKD enables two authenticated parties to establish a shared secret key in the form of a random bit string with information-theoretic security \cite{vajner2022quantum}. The resulting key can then be used for encryption schemes whose protection derives from the laws of quantum mechanics rather than computational complexity, ensuring immunity to any future advances in algorithms or processing power \cite{xu2020secure}.

The evolution of quantum cryptography began with the idea of conjugate coding proposed by Stephen Wiesner in the late 1960s and later formalized by Bennett and Brassard in the seminal BB84 protocol (1984) and Ekert’s E91 entanglement-based protocol (1991) \cite{wiesner1983conjugate,BB84,ekert1991quantum}. The BB84 scheme demonstrated secure key exchange using single photons and measurement-induced disturbance to reveal eavesdropping, while E91 employed entangled photon pairs to ensure security through quantum nonlocal correlations. These milestones established the foundation of QKD, where information is encoded in a \emph{two-dimensional Hilbert space} and each detected photon contributes at most one bit of secret key. In practice, however, channel attenuation and noise lower the detection rate and introduce errors, reducing the achievable secret key rate (SKR), which represents the effective key yield \cite{cerf2002security}. To overcome this limitation, researchers have explored \emph{higher-dimensional Hilbert spaces}, which increase the information per detected quantum and enhance the achievable SKR \cite{cerf2002security,bechmann2000quantum}.

High-dimensional QKD (HD-QKD) extends this idea by encoding information in a Hilbert space of dimension $d > 2$, allowing each photon to carry up to $\log_2 d$ bits \cite{cerf2002security}. Higher dimensionality not only increases the SKR but also improves error tolerance: for the same SKR, HD-QKD can tolerate higher quantum bit error rates (QBER) than conventional two-dimensional protocols \cite{shor2000simple}. This robustness is especially valuable in noisy \cite{patra2024dimensional}, imperfect real-world channels \cite{erhard2018twisted}. \textcolor{black}{For example, under security analyses against coherent attacks with one-way secret-key distillation, achieving a non-zero SKR requires that the QBER remains below 18.93\% for $d = 2$, while the corresponding sufficient security threshold increases to 24.70\% for $d = 4$ \cite{cerf2002security}, indicating the advantage of higher-dimensional encoding.}


The core purpose of QKD lies in its intrinsic ability to detect eavesdropping. Any attempt to intercept the key perturbs the transmitted quantum states and increases the QBER, as dictated by the information-disturbance principle \cite{fuchs1998information}. Once eavesdropping is detected, the corresponding data are discarded to preserve key secrecy. In HD-QKD, this disturbance grows with dimension $d$, making eavesdropping more readily detectable \cite{cerf2002security,portmann2022security}. Consequently, HD-QKD not only provides greater key capacity but also enhanced security, as eavesdropping-induced errors become more distinguishable, offering resilience against loss, noise, and channel attacks \cite{scarani2005quantum}.

In practice, the physical realization of HD-QKD relies on specific photonic degrees of freedom (DoFs), such as time-bin \cite{marcikic2002time,islam2017provably} and path encoding \cite{canas2017high,da2019stable}. Among these approaches, the orbital angular momentum (OAM) of light has emerged as a particularly powerful carrier \cite{spedalieri2006quantum}. OAM modes form an effectively unbounded basis \cite{allen1992orbital}, enabling in principle infinite-dimensional encoding \cite{erhard2018twisted}. Since the first OAM-based HD-QKD experiment in 2006 \cite{groblacher2006experimental}, extensive theoretical and experimental work has demonstrated OAM’s versatility and scalability \cite{mirhosseini2015high,larocque2017generalized,sit2017high,cozzolino2019orbital,wang2019characterizing,wang2021high,chang2023compact,sun2024experimental,scarfe2025fast,zhao2013large}. Building on this foundation, this survey consolidates current understanding of OAM-encoded HD-QKD, from its physical principles to experimental progress and emerging trends.

{\color{black}
\subsection{Literature Search Methodology} \label{subsec1.1}

This survey focuses on OAM-encoded HD-QKD, including representative implementations, system-level challenges in state generation, transmission, and detection, as well as recent developments in twin-field QKD (TF-QKD), continuous-variable QKD (CV-QKD), and measurement-device-independent or device-independent QKD (MDI/DI-QKD) architectures employing OAM.

The literature review was conducted using Google Scholar, IEEE Xplore, and Web of Science, with supplementary searches performed when necessary to improve coverage. Papers were retrieved using the keywords (OAM / Orbital Angular Momentum, Quantum Key Distribution / QKD, OAM-encoded QKD), including the combinations ``OAM / Orbital Angular Momentum and Quantum Key Distribution / QKD” and “OAM-encoded QKD."

The studies were selected according to the following criteria: peer-reviewed journal papers indexed in the Journal Citation Reports (JCR), published between January 2004 and March 12, 2026, directly related to OAM-encoded QKD or closely related high-dimensional quantum communication using OAM, and providing substantial theoretical, experimental, or system-level insights into protocol design, state generation, transmission, detection, turbulence effects, crosstalk, mitigation, or emerging directions.

The starting point of January 2004 was chosen based on early theoretical and experimental milestones \cite{spedalieri2006quantum, groblacher2006experimental}. The initial search identified 498 records, including 332 from Google Scholar, 6 from IEEE Xplore, 144 from Web of Science, and 16 from supplementary sources. After duplicate removal, 351 records remained, and 70 full-text articles were finally selected as the core reviewed set of this survey. The main reasons for exclusion were lack of direct relevance to OAM-encoded QKD, exclusive focus on classical OAM communication, or insufficient theoretical, experimental, or system-level relevance.

In addition to the core screened studies, a limited number of foundational references on conventional QKD, HD-QKD, mutually unbiased bases (MUBs), and OAM were included to provide necessary background, but were not counted as part of the core reviewed set.}

{
\color{black}

\setlength{\cmidrulewidth}{0.02em}
\begin{table*}[!t]
\captionsetup{font={color=black}}
\color{black}
\centering
\caption{Summary of related reviews on OAM-encoded QKD.}
\label{tab:survey_comparison}
\renewcommand{\arraystretch}{1.15}
\newcolumntype{M}[1]{>{\centering\arraybackslash}m{#1}}
\newcolumntype{L}[1]{>{\raggedright\arraybackslash}m{#1}}
\resizebox{\textwidth}{!}{
\begin{tabular}{L{2.5cm} M{1.2cm} L{2.8cm} M{1.5cm} M{1.5cm} M{1.8cm} M{1.8cm} M{2cm} L{5cm}}
\toprule
\textbf{Survey Reference} & \textbf{Year} & \textbf{Main Topic and Scope} & \textbf{Specifically About OAM?} & \textbf{Covers HD-QKD?} & \textbf{Includes Experimental Implementations?} & \textbf{Discusses System-Level Challenges?} & \textbf{Covers Emerging Directions (TF, CV, MDI, DI)?} & \textbf{Main Limitation Relative to Manuscript} \\
\midrule
\cite{erhard2018twisted} & 2018 & Quantum perspectives of twisted photons in high dimensions. & Yes & Yes & Yes & Yes (basic physics) & No & Emphasizes foundational physics and entanglement more than practical secure-network implementations and intelligent compensation techniques. \\
\cmidrule(lr){1-9}
\cite{cozzolino2019high} & 2019 & High-dimensional quantum communication: benefits, progress, and challenges. & No & Yes & Yes & Yes (fiber/free-space) & No & Published before the most recent advances; treats spatial modes together with path and time-bin encodings, and does not cover recent developments in adaptive optics and network architectures. \\
\cmidrule(lr){1-9}
\cite{otte2020high} & 2020 & High-dimensional cryptography with spatial modes of light. & No & Yes & Yes (basic) & No & No & Mainly serves as a tutorial on spatial-mode-based high-dimensional cryptography rather than a comprehensive review of modern implementation challenges. \\
\cmidrule(lr){1-9}
\cite{pirandola2020advances} & 2020 & Advances in comprehensive quantum cryptography and networks. & No & Yes (briefly) & Yes (extensive) & Yes (general QKD) & Yes (TF, CV, MDI, DI) & Overly broad; lacks any specific focus on spatial modes, state generation hardware, or specific channel physics like diffraction. \\
\cmidrule(lr){1-9}
\cite{willner2021orbital} & 2021 & OAM of light for optical communications. & Yes & Yes (briefly) & Yes (mostly classical) & Yes (classical optical) & No & Primarily focused on classical data links; lacks rigorous quantum security proofs, decoy-state analysis, and deep protocol evaluation. \\
\cmidrule(lr){1-9}
\cite{koudia2025physical} & 2025 & Physical-layer aspects of quantum communications, including MIMO and CV-related perspectives. & No & No & No & Yes (MIMO) & Yes (CV, MIMO) & Focuses on generalized spatial diversity and MIMO theory rather than detailed discrete-variable or OAM-based high-dimensional implementations. \\
\cmidrule(lr){1-9}
\textbf{This work} & 2026 & OAM-encoded HD-QKD: foundations, experiments, and recent trends. & Yes & Yes & Yes & Yes (exhaustive specific analysis) & Yes (TF, CV, MDI, DI) & -- \\
\bottomrule
\end{tabular}}
\end{table*}
}

{\color{black}
\subsection{Related Reviews and Our Contributions}

Existing review literature related to OAM-encoded HD-QKD is distributed across several partially overlapping perspectives. Prior surveys have addressed the physics of twisted photons and high-dimensional entanglement \cite{erhard2018twisted}, high-dimensional quantum communication frameworks \cite{cozzolino2019high}, tutorial-style introductions to spatial-mode cryptography \cite{otte2020high}, general quantum cryptography and quantum-network architectures \cite{pirandola2020advances}, classical OAM-based optical communication \cite{willner2021orbital}, and physical-layer perspectives on quantum communications \cite{koudia2025physical}. Although these works provide important background, a dedicated review focusing specifically on OAM-encoded HD-QKD remains lacking.

Erhard \emph{et al.} \cite{erhard2018twisted} discussed the quantum properties of twisted photons, including OAM states and high-dimensional entanglement, providing the underlying physical foundation. Cozzolino \emph{et al.} \cite{cozzolino2019high} surveyed high-dimensional quantum communication across multiple degrees of freedom, where OAM-encoded HD-QKD was treated as part of a broader framework. Otte \emph{et al.} \cite{otte2020high} presented a tutorial-oriented introduction to spatial-mode cryptography, focusing on basic principles rather than systematic comparisons of implementations. Pirandola \emph{et al.} \cite{pirandola2020advances} reviewed modern quantum cryptography and quantum networks, including advanced paradigms such as TF-QKD, CV-QKD, and MDI-/DI-QKD, but without detailed treatment of OAM-specific implementation issues. Willner \emph{et al.} \cite{willner2021orbital} focused on classical OAM-based optical communication, with only brief discussion of quantum links. More recently, Koudia \emph{et al.} \cite{koudia2025physical} considered physical-layer aspects of quantum communications, emphasizing channel modeling rather than discrete-variable OAM-based HD-QKD implementations.

A comparative summary of these reviews and the positioning of this work is provided in Tab.~\ref{tab:survey_comparison}. As indicated, despite extensive studies on OAM physics, high-dimensional quantum communication, and quantum-cryptographic architectures, there is no review that systematically addresses OAM-encoded HD-QKD from the combined perspectives of physical foundations, representative experiments, system-level bottlenecks, and emerging research directions. Accordingly, this survey provides a structured overview of OAM-encoded HD-QKD, including prepare--measure and entanglement-based implementations, practical challenges in state generation, transmission, and detection, and recent developments in advanced QKD architectures.
}

{\color{black}
\subsection{Structure of the Survey}

The overall structure of this survey is illustrated in Fig.~\ref{fig:structure}, which provides a visual overview of the organization of the paper. While the literature search, screening, and selection procedure are described in Section~\ref{subsec1.1}, the figure reflects the outcome of this process through the classification and analysis of the surveyed works. As shown in Fig.~\ref{fig:structure}, the studies are organized into three main categories—implementations, challenges, and emerging directions—corresponding to Sections~\ref{application}, \ref{challenges}, and \ref{future}, respectively.

The remainder of this paper is organized as follows. Section~\ref{qkd_background} provides background on OAM-encoded QKD, including the QKD workflow and post-processing (Section~\ref{sub_QKD}), HD-QKD security metrics (Section~\ref{sub_HD_QKD}), and OAM-based HD-QKD implementations (Section~\ref{sub_OAM_QKD}). Section~\ref{application} reviews representative experimental demonstrations under prepare--measure (Section~\ref{sub_Prepare}) and entanglement-based schemes (Section~\ref{sub_Ent}), followed by a benchmark-based comparison (Section~\ref{sec3_add}). Section~\ref{challenges} discusses practical limitations in state generation (Section~\ref{Generation}), transmission (Section~\ref{Transmission}), and reception (Section~\ref{Reception}). Section~\ref{future} surveys recent research directions, including TF-QKD (Section~\ref{sub_TF}), CV-QKD with spatial-mode multiplexing (Section~\ref{sub_CV}), and MDI/DI-QKD frameworks (Section~\ref{MDI}). Finally, Section~\ref{conc} concludes the paper.

}

{
\color{black}

\begin{figure}
\captionsetup{font={color=black}}
\color{black}
    \centering
    \includegraphics[width=1\linewidth]{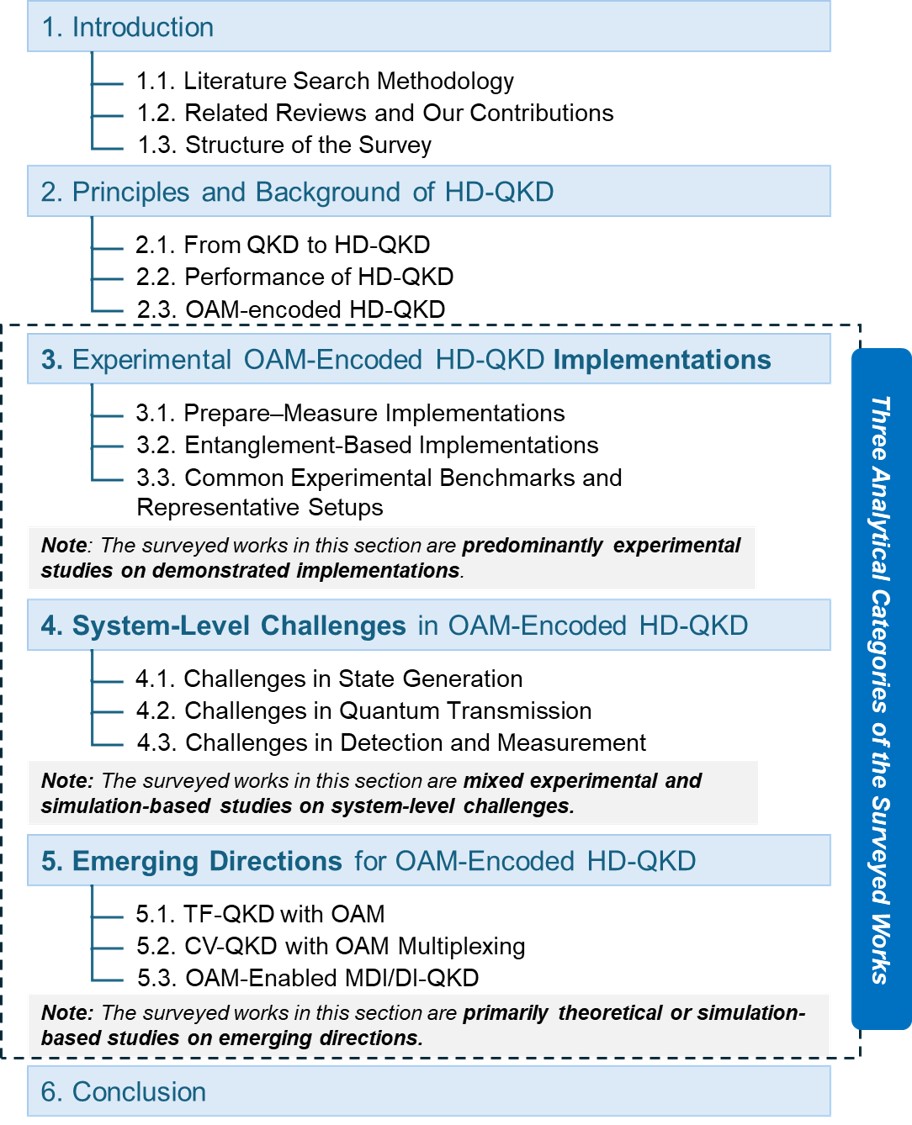}
    \caption{The outline of this survey paper.}
    \label{fig:structure}
\end{figure}

}


\section{Principles and Background of HD-QKD}\label{qkd_background}
\subsection{From QKD to HD-QKD}\label{sub_QKD}
The objective of QKD is to enable two authenticated parties to establish a shared symmetric secret key with information-theoretic security \cite{BB84}. 
In QKD, each transmitted photon represents a \emph{quantum state}, described mathematically as a normalized vector in a complex Hilbert space. The dimension of this space defines the number of distinguishable states, and information is encoded by selecting quantum states from different bases within this space.
In this paper, a qubit denotes a two-dimensional system. A qudit denotes a $d$-level system. When $d=3$ and $d=4$ we may use the specific terms qutrit and ququart, respectively. Unless stated otherwise, qudit will mean $d>2$ and we use HD-QKD to refer to QKD using qudits. QKD protocols are commonly classified into prepare–measure and entanglement-based schemes (e.g., BB84 \cite{BB84} and E91 \cite{ekert1991quantum}, respectively). Throughout this section we adopt the standard convention that Alice is the sender, Bob is the receiver, and Eve denotes a potential eavesdropper.

Many QKD protocols, including both prepare–measure and entanglement-based schemes, follow a common working process \cite{vajner2022quantum}.

(1) \textbf{Qubit exchange:} In the prepare–measure scheme, Alice encodes single-photon states in one of at least two randomly chosen, non-commuting bases and sends them to Bob through a quantum channel, where Bob measures each photon using a randomly selected basis on his side. Such prepare–measure schemes that use two non-commuting measurement bases are commonly referred to as two-basis protocols, since their security arises from the incompatibility of the two measurement bases. In practice, the two bases are typically chosen to be MUBs, with the standard BB84 protocol being a canonical example. In the entanglement-based scheme, an entangled photon source generates pairs and distributes one photon to Alice and the other to Bob, after which both parties perform local measurements with randomly chosen bases. The time-ordered sequences recorded from these quantum measurements, before any public discussion or post-processing, form the raw keys held by Alice and Bob. Because basis choices often differ and the quantum channel and devices introduce errors, Alice’s and Bob’s raw keys are generally not identical, only correlated.

(2) \textbf{Sifting:} Over an authenticated classical channel, Alice and Bob reveal only the bases they used for each transmission and keep the trials where their basis choices match. The actual bit values remain undisclosed. They then retain the corresponding entries in their raw keys, forming the \emph{sifted keys}. Under an idealized condition of a noiseless channel and no eavesdropping, these sifted keys would be perfectly identical for Alice and Bob.

(3) \textbf{Parameter estimation:} Even after sifting, discrepancies remain between Alice’s and Bob’s sifted keys because of channel impairments, device imperfections, or potential eavesdropping from Eve. To quantify these disturbances, they publicly reveal a randomly chosen subset of the sifted key over the classical channel and compare the disclosed entries. In the case of eavesdropping, the \emph{information-disturbance principle} of quantum mechanics \cite{fuchs1998information} states that any attempt to gain information about the key, including a measurement in a basis incompatible with the encoding, necessarily perturbs the transmitted states \cite{portmann2022security}. Such disturbances appear as mismatches between Alice and Bob even when they used the same basis, and these mismatches are measured by the QBER. In practice, QBER is estimated from the revealed subset, and finite-size statistical bounds are applied to infer an upper limit on the error rate of the unrevealed portion. The protocol continues only if the observed QBER lies below the threshold required for a non-zero SKR; otherwise, the process is aborted.

(4) \textbf{Information reconciliation:} After parameter estimation, Alice and Bob perform information reconciliation over an authenticated public classical channel to align their sifted keys. During this process, they iteratively exchange parity information, revealing only block parities rather than actual key symbols, to locate and correct mismatched bits. Unlike conventional error correction in classical communication, this exchange is interactive and publicly observable, so the disclosed parity data are conservatively regarded as information available to Eve and are excluded from the final key.

(5) \textbf{Privacy amplification:} After information reconciliation, Alice and Bob shorten the reconciled key to remove any information that might be available to Eve. This compression is typically achieved using randomly chosen universal hash functions, which ensure that Eve’s knowledge about the final key becomes negligible. The remaining bits constitute the final secret key. A key confirmation step is then performed to verify that both parties hold identical keys. The SKR is defined as the number of secure bits generated per unit time, typically measured in bits per second.

\begin{figure*}[t]
    \centering
    \includegraphics[width=1\linewidth]{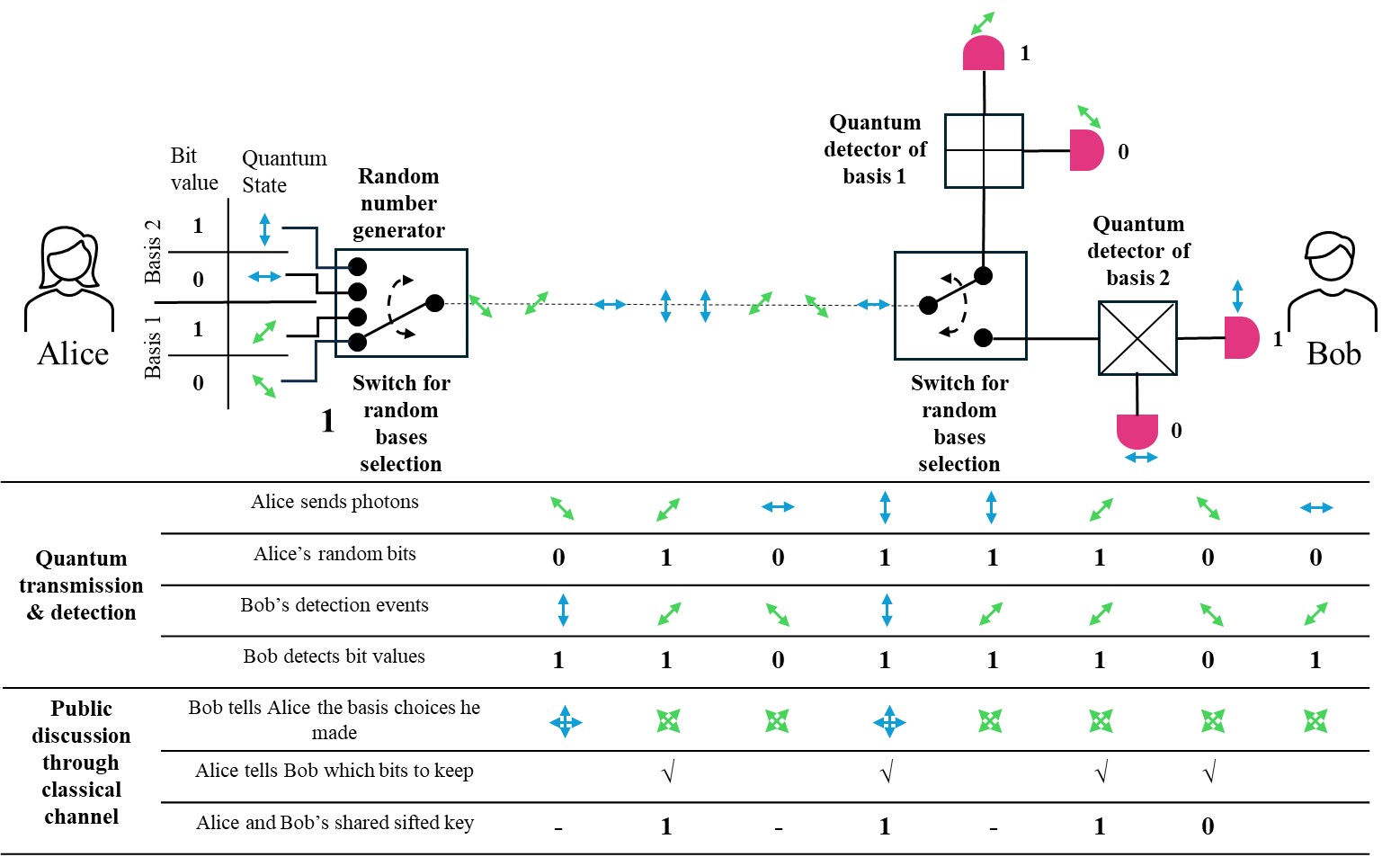}
    \caption{Idealized BB84 workflow showing qubit exchange (top) and sifting (bottom) under noiseless, lossless conditions without eavesdropping.}
    \label{fig:BB84_qkd}
\end{figure*}

To illustrate the process, Fig.~\ref{fig:BB84_qkd} shows an idealized BB84 workflow based on qubits and two MUBs, representing the standard form of a two-basis protocol. In a prepare–measure setting, Alice encodes each photon in one of the two bases chosen at random, and Bob measures each received photon using a randomly chosen basis as well.  
When they select the same basis, Bob correctly recovers Alice’s bit. When they choose different bases in this ideal scenario, Bob’s outcome is completely random between $0$ and $1$, each occurring with probability $1/2$, so it reveals no information about Alice’s choice. This outcome reflects the defining property of MUBs, in which a state from one basis has equal overlap with all states of the other. The illustration assumes no eavesdropping from Eve, a noiseless and lossless quantum channel, and ideal devices. In the lower part of the figure, each column represents one transmission round through the quantum channel. Rounds with matching bases are retained to form the sifted key, while rounds with mismatched bases are discarded. Under these idealized conditions, the parameter estimation step would yield $\mathrm{QBER}=0$. Information reconciliation then requires no additional communication cost, and privacy amplification is unnecessary beyond authentication and key confirmation. The sifted key therefore becomes the final secret key, and the figure highlights only the first two stages of the protocol: qubit exchange and sifting.

The workflow above yields a shared symmetric secret key between Alice and Bob. To achieve confidentiality guaranteed by information theory, this key can be used with a one-time pad (OTP) to encrypt messages, which remains the only provably secure encryption method. The OTP, first proposed by Vernam in 1926 \cite{vernam1926cipher}, requires that (1) the key shared by Alice and Bob is secret, (2) it is randomly generated, (3) its length is at least equal to that of the message, and (4) it is used only once. Among these conditions, maintaining secrecy is the most difficult to ensure through classical key establishment, which at best provides computational security \cite{vajner2022quantum}. In contrast, QKD achieves information-theoretic secrecy by linking any eavesdropper’s knowledge to measurable disturbance and removing it through privacy amplification. Using keys generated by QKD with a properly managed OTP satisfies all four conditions and enables truly end-to-end confidential communication \cite{scarani2009security,sheridan2010security,xu2020secure}. 

The workflow above shows that SKR serves as a primary performance indicator for QKD, directly reflecting the amount of secure information generated per transmitted photon. Improving SKR is therefore a central goal in system design. One effective approach is to extend beyond the two-dimensional encoding of conventional QKD and employ higher-dimensional systems, where each photon carries more than one bit of information. Such high-dimensional encoding enhances information efficiency and increases robustness to noise and imperfections. However, moving to higher dimensions also introduces additional considerations in the choice of encoding bases and overall system complexity, which are discussed in the next section.

\subsection{Performance of HD-QKD}\label{sub_HD_QKD}
As described by Sheridan \emph{et~al.} \cite{sheridan2010security}, a $d$-dimensional BB84 protocol can be formulated using two MUBs. Each basis contains $d$ orthogonal states, and Alice and Bob randomly select between the two bases for each transmission. The sifted key is formed from the rounds in which their basis choices coincide.

In this model, the quantity $Q$ represents the probability that Alice and Bob obtain different measurement outcomes when they happen to use the same basis for key generation.
This probability is referred to as the \emph{symbol-error probability}, since each transmission round yields a pair of measurement outcomes—one from Alice and one from Bob—forming a single symbol.
Among all transmission rounds, only those in which the two parties select the same basis constitute the \emph{sifted key}, on which $Q$ is evaluated.
Assuming symmetric noise as in Sheridan \emph{et~al.} \cite{sheridan2010security}, that is, a $d$-dimensional generalization of depolarizing noise with identical statistics across both bases, the sifted key can be modeled as a $d$-ary symmetric channel with crossover probability $Q$.  In an idealized regime where a large number of signals are exchanged and only one-way classical communication is used for post-processing, the extractable \emph{secret key fraction} per sifted symbol is expressed as
\begin{equation}
r_d(Q) = \log_2 d - H(A|B) - I_E .
\end{equation}
This quantity represents the fraction of sifted data that can be converted into final secure key bits. Here, $H(A|B)$ denotes the amount of information revealed during reconciliation—the minimum number of bits that must be publicly disclosed for Bob’s string to match Alice’s—while $I_E$ represents Eve’s \emph{Holevo information}, the maximum information that a quantum eavesdropper could, in principle, obtain about Alice’s raw $d$-ary symbol.
For the sifted key, the communication between Alice and Bob can be modeled as a $d$-ary symmetric channel, for which the reconciliation information equals the conditional entropy,
\begin{equation}
H(A|B)=h_d(Q),
\end{equation}
where
\begin{equation}
h_d(Q) = - (1-Q)\log_2(1-Q) - Q \log_2 \Big(\frac{Q}{d-1}\Big),
\end{equation}
representing the uncertainty of Bob’s outcome given Alice’s symbol under the observed error probability $Q$.
In the two-basis protocol analyzed by Sheridan \emph{et~al.} \cite{sheridan2010security}, the symmetry of this $d$-ary channel implies that Eve’s Holevo information takes the same form,
\begin{equation}
I_E = h_d(Q).
\end{equation}
Substituting these relations yields
\begin{equation} \label{for_thereshold}
r_d(Q) = \log_2 d - 2 h_d(Q).
\end{equation}
Analytical expressions for \eqref{for_thereshold} are generally unavailable, and numerical threshold values reported in Sheridan \emph{et al.} \cite{sheridan2010security} and subsequent studies are widely cited in the literature. It is also observed that $Q_{\mathrm{th}}(d)$ increases monotonically with dimension $d$, illustrating the enhanced error tolerance of higher-dimensional protocols.

To illustrate how the dimension-dependent thresholds $Q_{\mathrm{th}}(d)$ translate into practical transmission limits, we consider a simple link–noise model widely used in QKD research. This model relates the symbol-error probability $Q$ to physical parameters of the channel and receiver, including two dominant noise sources: channel loss and detector dark counts, the latter referring to detection events that occur without a received photon at Bob’s receiver.  In a $d$-dimensional prepare–measure system, Bob employs $d$ parallel single-photon detectors, each corresponding to one of the $d$ possible outcomes in the key basis. Ideally, when Alice transmits a photon representing one symbol, only the correct detector at Bob registers a detection event. Channel loss can cause the photon to be absorbed or scattered before it reaches Bob, leading to no photon detection during that transmission round. Dark counts, on the other hand, produce false detections randomly distributed among the $d$ detectors, resulting in uniformly random outcomes. Throughout this discussion, we assume a low detection-probability regime, where at most one detection event occurs per transmission round, consistent with the typical assumptions in QKD link modeling.

As a representative case, we consider a fiber-based quantum channel, where transmission loss is dominated by standard fiber attenuation.
For a typical single-mode fiber with attenuation $\alpha = 0.2~\mathrm{dB/km^{-1}}$ \cite{cozzolino2019high}, the transmittance after a fiber length $L$ (in km) is
\begin{equation}
T(L) = 10^{-\alpha L/10}.
\end{equation}
In addition to fiber loss, the overall transmission efficiency also depends on the optical coupling and detector performance, which we represent by a system efficiency factor $\eta$.
Under these conditions, the per-gate probability that a signal photon produces a detection event is
\begin{equation}
P_{\mathrm{sig}}(L) = \eta T(L).
\end{equation}
Let $p_d \ll 1$ denote the dark-count probability of a single detector per transmission round.
In a $d$-dimensional receiver, Bob operates $d$ independent detectors, each corresponding to one symbol in the key basis.
The total probability that a detection event arises from a dark count across all detectors is then
\begin{equation}
P_{\mathrm{dc}} = d p_d.
\end{equation}
As formulated by Scarani \emph{et~al.} \cite{scarani2009security}, if the fraction of detections originating from actual signal photons is defined as
\begin{equation}
Y(L) = \frac{P_{\mathrm{sig}}(L)}{P_{\mathrm{sig}}(L) + P_{\mathrm{dc}}},
\end{equation}
then $1-Y(L)$ represents the fraction due to dark counts.
Dark counts are generally considered basis-independent in QKD modeling \cite{scarani2009security} and therefore yield uniformly random outcomes within the $d$-ary alphabet.
In the qubit case, this corresponds to a random-error probability of $\frac{1}{2}$, whereas for dimension $d$ the corresponding value is $\frac{d-1}{d}$.
Consequently, the resulting symbol-error probability including dark counts is
\begin{equation}
Q_{\mathrm{dc}}(L) = \bigl(1-Y(L)\bigr) \frac{d-1}{d}.
\end{equation}
Under the asymptotic assumption with one-way post-processing, and considering only dark counts as the error source, the remaining error tolerance that can be sustained at distance $L$ while still yielding a positive key fraction is
\begin{equation} 
Q_{\max}(L)=\max\!\big\{0,\;Q_{\mathrm{th}}(d)-Q_{\mathrm{dc}}(L)\big\}. 
\end{equation}

\begin{figure}
    \centering
    \includegraphics[width=1\linewidth]{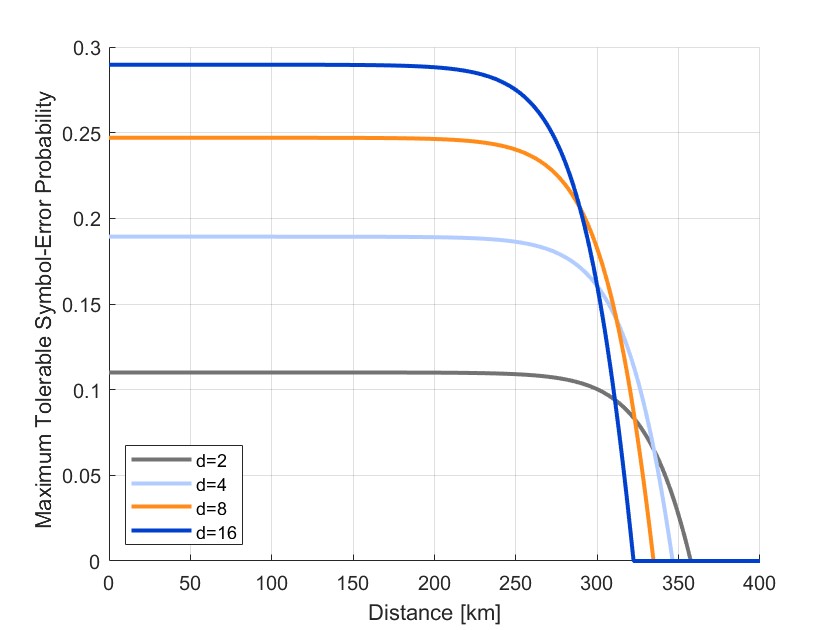}
    \caption{Maximum tolerable symbol-error rate $Q_{\max}(L)$ as a function of fiber distance $L$ for $d = 2, 4, 8,$ and $16$ in a $d$-dimensional BB84 protocol using two MUBs under idealized security assumptions.}
    \label{qber_dis}
\end{figure}

Fig.~\ref{qber_dis} shows the remaining tolerable error $Q_{\max}(L)$ as a function of fiber length $L$ for several dimensions $d$.
Each curve represents an ideal single-photon $d$-dimensional BB84 protocol using two MUBs under standard theoretical assumptions and a channel model that includes only fiber loss and detector dark counts \cite{BB84,sheridan2010security,scarani2009security,cozzolino2019high}.
For short transmission distances where $L\approx0$, the tolerable error approaches $Q_{\mathrm{th}}(d)$, which increases with $d$ and reflects the higher intrinsic error tolerance of high-dimensional protocols.
As the distance increases, the signal fraction $Y(L)$ decreases, $Q_{\mathrm{dc}}(L)$ grows, and $Q_{\max}(L)$ drops more rapidly for larger $d$ because both $P_{\mathrm{dc}} = d,p_d$ and the random-error factor $(d-1)/d$ scale with dimension.
Consequently, high-dimensional QKD protocols exhibit stronger noise resilience at short ranges, while two-dimensional qubit systems achieve the longest transmission distances.
In practice, however, the overall performance depends on the specific hardware and operating conditions, and there are scenarios in which high-dimensional protocols can outperform qubits even in achievable distance \cite{cozzolino2019orbital,wang2021high}.

\subsection{OAM-encoded HD-QKD}\label{sub_OAM_QKD}

To physically implement the high-dimensional encoding discussed in the previous section, OAM modes of light provide a natural and widely used platform.
OAM describes the helical phase structure of an optical field, characterized by the azimuthal phase term $\exp(il\theta)$,
where $l$ is the azimuthal index determining the number of phase windings around the propagation axis
and $\theta$ denotes the azimuthal coordinate in the plane transverse to the propagation direction.
Each OAM mode may also exhibit variations in its radial intensity, described by the radial index $p$. The azimuthal and radial indices are conveniently represented by Laguerre–Gaussian (LG) functions \cite{allen1992orbital}, which form an orthogonal basis encompassing the azimuthal and radial degrees of freedom.
The complex amplitude of an OAM mode labeled by $(l,p)$ is given by
\begin{align}\label{LG}
\varphi_{l,p}(r,\theta,z)=&
\frac{A}{w(z)}
\left(\frac{\sqrt{2}\,r}{w(z)}\right)^{|l|}
L_{p}^{|l|}\!\left(\frac{2r^{2}}{w^{2}(z)}\right)\notag\\ 
&\times \exp\!\left(-\frac{r^{2}}{w^{2}(z)}\right)
\exp(il\theta),
\end{align}
where  
$A = \sqrt{\frac{2p!}{\pi\, (|l|+p)!}}$ is the normalization factor.   
The beam radius along propagation is  
$w(z) = w_0 \sqrt{1 + \frac{z^2}{z_R^2}},$  
where $w_0$ is the beam waist and  
$z_R = \frac{1}{2}k w_0^2,$  
with $k = 2\pi/\lambda$ and $\lambda$ being the wavelength.  
$L_{p}^{|l|}(\cdot)$ denotes the generalized Laguerre polynomial, and $(r,\theta,z)$ are cylindrical coordinates.  
Any two OAM modes with different $l$ or $p$ are orthogonal under the transverse inner product.  
In typical OAM-based QKD implementations \cite{erhard2018twisted}, the radial index is fixed at $p = 0$,  
so the mode number is determined solely by $l$, with $+l$ and $-l$ treated as distinct orthogonal modes.

Because OAM modes possess an intrinsically high-dimensional structure, they provide a natural physical platform for implementing HD-QKD.  
Their helical phase profile allows a large number of mutually orthogonal spatial modes, and the accessible OAM values can in principle span an unbounded Hilbert space \cite{molina2001management}.  
These properties have made OAM one of the most widely explored spatial degrees of freedom for high-dimensional quantum communication \cite{zhang2023orbital,mirhosseini2015high,larocque2017generalized,sit2017high,cozzolino2019orbital,wang2019characterizing,wang2021high,wang2020high,chang2023compact}.

\begin{figure}
    \centering
    \includegraphics[width=1\linewidth]{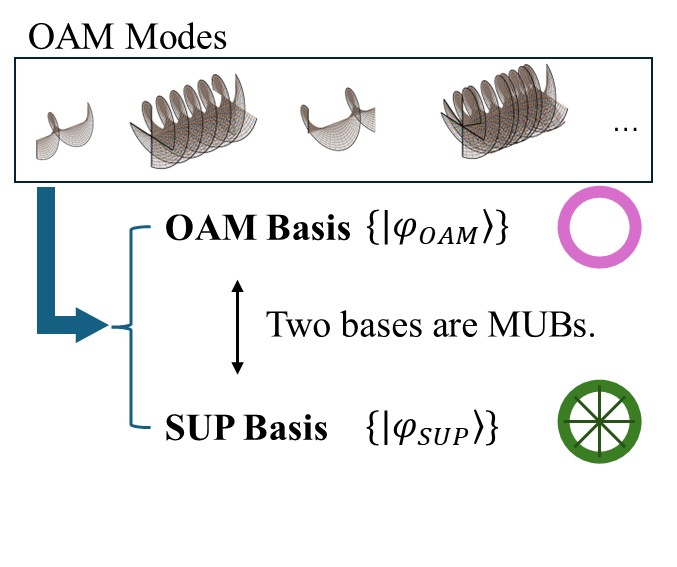}
    \caption{OAM modes provide a big alphabet, used in two MUBs.}
    \label{fig:MUBS}
\end{figure}

The high-dimensional structure of OAM modes naturally leads to a convenient choice of encoding bases for HD-QKD.  In OAM-encoded HD-QKD, the OAM basis and the superposition (SUP) basis form two MUBs, which together constitute a two-basis protocol.
The OAM modes are defined for $l\in[-L,L]$, so the protocol operates in a $d$-dimensional space with $d=2L+1$.  
The OAM basis consists of distinct OAM modes represented as
\begin{equation}
\{\,\ket{\varphi_{\mathrm{OAM}}}\,\}
= \{\ket{-L},\ \ket{-L+1},\ \ldots,\ \ket{0},\ \ldots,\ \ket{L-1},\ \ket{L}\}.
\end{equation}
The SUP basis is an equal superposition of OAM modes with a fixed relative phase between adjacent components \cite{erhard2018twisted}:
\begin{equation}
\{\,\ket{\psi_{\mathrm{SUP}}}\,\}
= \left\{\frac{1}{\sqrt{d}}
\sum_{l=-L}^{L} \exp\!\left(i\frac{jl\,2\pi}{d}\right)\ket{\varphi_{\mathrm{OAM}}}\right\}.
\end{equation}
As summarized in Fig.~\ref{fig:MUBS}, OAM modes provide a $d$-dimensional alphabet $\{\ket{l}\}_{l=-L}^{L}$ with $d=2L+1$, employed in two MUBs: the OAM basis and its SUP basis.

{
\captionsetup{font={color=black}}
\captionsetup[subfigure]{font={color=black}}
\begin{figure*}[t]
  \centering
  \begin{subfigure}[t]{0.48\linewidth}
    \centering
    \includegraphics[width=\linewidth]{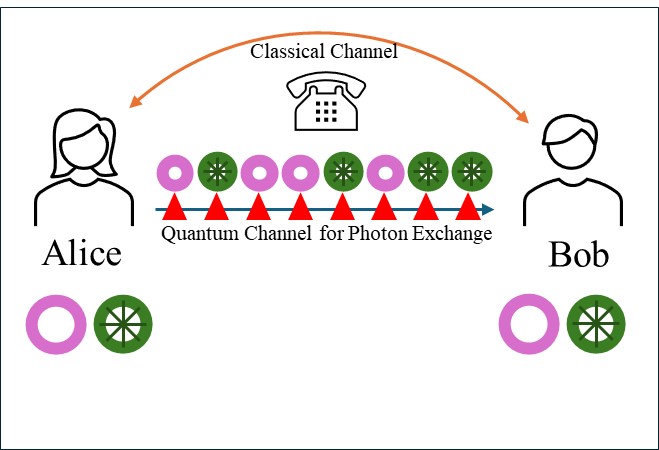}
    \caption{Prepare–measure QKD}
    \label{fig:oam-pre}
  \end{subfigure}\hfill
  \begin{subfigure}[t]{0.48\linewidth}
    \centering
    \includegraphics[width=\linewidth]{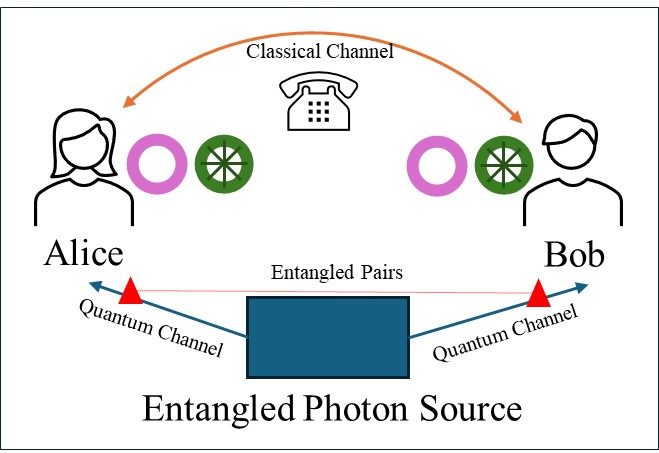}
    \caption{Entanglement-based QKD}
    \label{fig:oam-ent}
  \end{subfigure}
  \caption{OAM-encoded HD-QKD of two schemes (a) prepare–measure, (b) entanglement-based.}
  \label{fig:oam-two}
\end{figure*}
}

For clarity of implementation, Fig.~\ref{fig:oam-two} summarizes the two architectures employed in OAM-encoded HD-QKD. In the prepare–measure setting, Alice selects either the OAM or SUP basis at random and prepares the corresponding state, while Bob performs a measurement in a randomly chosen basis. The classical channel is then used for sifting and one-way post-processing.  In the entanglement-based architecture, a source distributes OAM-entangled photon pairs to Alice and Bob, who independently choose between the OAM and SUP bases for their local measurements before classical post-processing.  
In both implementations, the same two MUBs realize a $d$-dimensional two-basis protocol, with the usable alphabet size determined by the choice of $L$.

\section{Experimental OAM-Encoded HD-QKD Implementations}\label{application}
\textcolor{black}{This section reviews experimentally demonstrated implementations of OAM-encoded HD-QKD, with a focus on establishing a clear taxonomy of the existing experimental landscape. The discussion is organized by protocol type, namely prepare–measure schemes in Section~\ref{sub_Prepare} and entanglement-based schemes in Section~\ref{sub_Ent}, while each experiment is further categorized by channel type (free-space or fiber). This hierarchical classification allows different experimental approaches to be systematically compared and their relationships to be more clearly understood. Representative studies under these schemes are summarized in Tabs.~\ref{pam_qkd} and~\ref{en_qkd}.
To complement this protocol-based classification, Section~\ref{sec3_add} and Tab.~\ref{tab:oam_hdqkd_comparison} provide a benchmark-oriented comparison across experimental conditions, system completeness, representative setups, SKR interpretation, and primary evaluation focus.
}


\subsection{Prepare–Measure Implementations}\label{sub_Prepare}
The first prepare–measure demonstration of OAM-encoded HD-QKD was reported by 
Mirhosseini \emph{et al.} (2015)~\cite{mirhosseini2015high}, who implemented a BB84-type 
protocol in a seven-dimensional OAM subspace ($d = 7$, $l \in \{-3, \dots, 3\}$). Their setup 
employed a digital micro-mirror device for state preparation and a refractive log--polar, 
fan-out--assisted mode sorter for detection. A mode sorter is an optical device that separates 
different spatial modes into distinct output positions so that each mode can be measured 
individually. Using this configuration, the experiment achieved approximately $93\%$ separation  efficiency, about $2.05$ bit/sifted photon, and a symbol-error probability near $10.5\%$. Building on this foundation, Larocque \emph{et al.} (2017)~\cite{larocque2017generalized} developed a liquid-crystal Pancharatnam--Berry optical element (PBOE) that functions as a high-performance mode sorter for OAM modes. With this device, the experiment realized a $10$-dimensional state space (corresponding to $d = 10$), constructed as the tensor product of  five OAM modes and two polarization states. The implementation exhibited a QBER of 
$e_b \approx 21.2\%$ and a modal confinement of about $72\%$, meaning that only $72\%$ of the 
optical power was detected in the intended output channel, with the remainder leaking into 
neighboring channels due to optical aberrations. The authors also identified clear upgrade 
directions by proposing refined fan-out holograms, which improve mode separation, together 
with enhanced PBOE fabrication techniques. These improvements are expected to reduce 
crosstalk and increase the overall system-level SKR.

While the earlier demonstrations primarily examined device-level mode sorting and the joint 
measurement of polarization and OAM, channel robustness in a realistic outdoor link was 
investigated by Sit \emph{et al.} (2017)~\cite{sit2017high}. They implemented a $d = 4$ hybrid 
polarization-OAM BB84 link over a $300$ m urban free-space channel using a heralded 
single-photon source. Instead of log--polar mode sorting, the experiment relied on q-plates for state preparation and on polarization-based measurements for analysis, providing a compact and stable alternative to mode-sorting architectures. The reported QBER was $11\%$, and the 
information yield reached $0.65$ bit/sifted photon. Both quantities lie within the $d = 4$ 
security threshold, showing that hybrid polarization--OAM encoding can preserve security under 
moderate atmospheric turbulence and demonstrating the feasibility of extending OAM-based QKD 
beyond controlled laboratory environments.

Rather than improving turbulence robustness through hybrid polarization--OAM encoding, 
Scarfe \emph{et al.} (2025)~\cite{scarfe2025fast} investigated receiver-side compensation by 
introducing fast adaptive optics (AOs) at the detection stage of a free-space prepare–measure 
HD-BB84 link. Their experiment used OAM and SUP as MUBs and emulated turbulence over an 
effective propagation path of approximately $70$ cm in the laboratory. The study quantified 
QBER with the AO system enabled and disabled for even dimensions $d \in \{2,4,6,8,10\}$. 
With AO enabled, the sifted-key rates reached $0.851$ bit/sifted photon for $d = 2$, 
$1.06$ bit/sifted photon for $d = 4$, $0.991$ bit/sifted photon for $d = 6$, and 
$0.412$ bit/sifted photon for $d = 8$. The case $d = 10$ did not yield a secure rate because 
the SUP measurements exceeded the theoretical QBER threshold. These results illustrate a 
practical resolution limit for SUP measurements at higher dimensions and demonstrate that 
high-speed, high-resolution AO using a $97$-actuator deformable mirror can transform a 
turbulence-degraded link into a secure prepare–measure OAM-encoded QKD channel.

Beyond free-space demonstrations, subsequent work shifted toward guided transmission and 
optical simplification. Cozzolino \emph{et al.} (2019)~\cite{cozzolino2019orbital} achieved 
$1.2$ km C-band transmission through an air-core fiber by preparing four OAM modes 
$l \in \{-7, -6, 6, 7\}$ using vortex plates and spin-orbit conversion, followed by detection 
with a Mach--Zehnder and Dove-prism mode sorter. Their implementation also incorporated 
decoy-state weak coherent pulses, aligning the experiment with contemporary QKD practice. 
Complementing this fiber-oriented direction, Wang \emph{et al.} (2019) \cite{wang2019characterizing} used vector--vortex nonseparability to map $d = 4$ OAM state preparation and measurement onto polarization-domain unitaries and analyzers. This approach yielded a compact and phase-stable implementation with a QBER of $0.60\% \pm 0.06\%$ and 
$1.849 \pm 0.008$ bit/sifted photon.  In a related effort to reduce optical complexity while preserving the benefits of high  dimensionality, the mutually partially unbiased bases (MPUBs) framework introduced basis switching between same-order LG and Hermite--Gaussian (HG) modes using only linear optics \cite{wang2020high}. A subsequent experiment validated this concept through a compact free-space implementation across dimensions $d \in \{2,4,5,6\}$, reporting well-characterized error rates and SKRs~\cite{chang2023compact}. Together, these results provide a practical and reproducible path from conceptual designs to deployable HD-QKD systems.

To extend transmission distance while maintaining a $d = 4$ alphabet, 
Wang \emph{et al.} (2021)~\cite{wang2021high} encoded four symbols in hybrid 
polarization-OAM states, combining two OAM modes with two polarization states.  
Using this encoding, they demonstrated secure operation over ring-core fiber links ranging 
from $4$ km to $25$ km, with QBER values remaining below the $d = 4$ security threshold.  
These results show that OAM-encoded HD-QKD can be adapted to metro-scale fiber distances 
without requiring substantial optical complexity. Focusing on implementation simplicity and reproducibility,  Sun \emph{et al.} (2024)~\cite{sun2024experimental} constructed a $d = 4$ state space from two polarization states and a two-mode OAM subspace, employed passive basis selection using forked-grating analysis, and implemented a decoy-state HD-BB84 protocol.  
Their system achieved $1.458$ bit/sifted photon, demonstrating that high-dimensional 
advantages can be realized with simplified optical components and offering practical guidance 
for compact and deployable HD-QKD designs.

Representative prepare–measure experiments are summarized in Tab.~\ref{pam_qkd}, which lists the channel type, year, transmission distance, protocol, dimensionality, and the reported key-rate metric for each demonstration. The \emph{Key rate} column follows the conventions of the original papers.  Values reported in \emph{bit/sifted photon} represent sifted-key rates, whereas values reported in \emph{bit/s} or \emph{bit/photon} correspond to SKRs as specified in the cited papers. \textcolor{black}{We retain the original SKR conventions reported in the cited works (e.g., bit/s, bit/photon, bit/sifted photon), because these quantities are not always directly interchangeable. In many OAM-encoded QKD demonstrations, the reported SKR values are theoretically inferred from experimentally measured error statistics or correlations under different security assumptions \cite{cerf2002security, shor2000simple, lo2005decoy, bouchard2018experimental, islam2017provably} and normalization conventions, rather than derived under a uniform SKR definition.}

\begin{table*}[!t]
\centering
\caption{Summary of OAM-encoded HD-QKD experiments in the prepare–measure scheme.}
\setlength{\tabcolsep}{6pt}
\begin{tabularx}{\textwidth}{@{}l c c l c c X@{}} 
\toprule
Channel & Year & Distance (km) & Protocol & Dimension & Key rate & Reference \\
\midrule
Free-space & 2015 & $ \approx 0$  & HD-BB84 & 7 & 6.5 bit/s & \cite{mirhosseini2015high} \\
Free-space & 2017 & $ \approx 0$ & HD-BB84 & 10 & 0.49 bit/photon &\cite{larocque2017generalized} \\
Free-space & 2017 & $0.3$ & HD-BB84 & 4 & 0.65 bit/sifted photon &\cite{sit2017high} \\
Hollow optical fiber & 2019 & $1.2$ & HD-BB84 & 4 & 42.3 kbit/s &\cite{cozzolino2019orbital} \\
Free-space & 2019 & $ \approx 0$ & HD-BB84 & 4 & 1.849 bit/sifted photon &\cite{wang2019characterizing} \\
Ring-core fiber & 2021 & $ 25$ & HD-BB84 & 4 & 0.201 bit/photon &\cite{wang2021high} \\
Free-space & 2023 & $ \approx 0$ & MPUBs & 6 & 1.4092 bit/sifted photon &\cite{chang2023compact} \\
Free-space & 2024 & $ \approx 0$ & HD-BB84 & 4 & 1.458 bit/sifted photon &\cite{sun2024experimental} \\
Free-space & 2025 & $ \approx 0$ & HD-BB84 & 8 & 0.412 bit/sifted photon &\cite{scarfe2025fast} \\
\bottomrule
\end{tabularx}
\label{pam_qkd}
\end{table*}

\begin{table*}[!t]
\centering
\caption{Summary of OAM-encoded HD-QKD experiments in the entanglement-based scheme.}
\setlength{\tabcolsep}{6pt}
\begin{tabularx}{\textwidth}{@{}l c c l c c X@{}} 
\toprule
Channel & Year & Distance (km) & Protocol & Dimension & Key rate & Reference \\
\midrule
Free-space & 2006 & $ \approx 0$  & E91 & 3 & - & \cite{groblacher2006experimental} \\
Free-space & 2013 & $ \approx 0$  &  \makecell[l]{Large-alphabet\\OAM-encoded QKD} & 2 & - & \cite{zhao2013large} \\
\bottomrule
\end{tabularx}
\label{en_qkd}
\end{table*}

\subsection{Entanglement-Based Implementations} \label{sub_Ent}

The role of OAM as a useful resource for photonic entanglement was first demonstrated by Mair \emph{et al.} (2001)~\cite{mair2001entanglement}, who generated OAM-entangled photon pairs across a five-dimensional space ($l \in \{-2, -1, 0, 1, 2\}$). This result established a foundation for high-dimensional entanglement and opened a pathway toward entanglement-based QKD. Building on this progress, the earliest OAM-encoded HD-QKD using the E91 protocol was 
reported by Gröblacher \emph{et al.} (2006)~\cite{groblacher2006experimental}, who realized qutrit pairs ($d = 3$) and demonstrated security through a three-dimensional Bell-inequality violation. Their experiment produced secret keys with an observed error rate of about $10\%$, providing a clear demonstration that high-dimensional OAM entanglement can support nonlocal security guarantees. 

Subsequent work explored alternative measurement strategies. Zhao \emph{et al.} (2013) \cite{zhao2013large} accumulated several entangled photon pairs within each time slot and used two-photon coincidence fringes to distinguish among $N$ non-orthogonal settings, enabling symbol decisions directly from coincidence measurements. In most OAM-encoded QKD schemes \cite{mirhosseini2015high, cozzolino2019orbital, scarfe2025fast, groblacher2006experimental}, different key symbols were mapped directly onto distinct orthogonal OAM eigenmodes, so that the size of the QKD alphabet closely matched the number of orthogonal OAM modes employed. By contrast, the authors altered the coding strategy itself: rather than enlarging the alphabet by introducing more orthogonal OAM eigenmodes, they realized a larger symbol alphabet through coincidence-based discrimination of multiple non-orthogonal sector-state settings. Although each photon effectively occupied a two-dimensional OAM subspace, this approach broadened the usable alphabet and clarified the trade-offs among source brightness, measurement stability, and decision reliability.

Researchers also investigated the highest OAM order that could be entangled reliably. Fickler \emph{et al.} (2016)~\cite{fickler2016quantum} generated polarization-OAM hybrid entanglement in which one photon was prepared in the states $\{\ket{+l}, \ket{-l}\}$ and verified non-separability up to $\lvert l \rvert = 10{,}010$. Although the encoding remained effectively  two-dimensional, the result highlighted the substantial headroom available for future high-dimensional entanglement sources and for mode-sorting technologies capable of resolving very large OAM indices.

A concise summary of representative entanglement-based OAM-encoded HD-QKD demonstrations is provided in Tab.~\ref{en_qkd}, which lists the channel type, year, transmission distance, protocol, dimensionality, and reported key-rate metric. A dash in the table indicates that an SKR was not reported in the original paper.  

Progress in this direction has been slower than in prepare–measure systems, largely because entanglement-based implementations require more demanding hardware. Practical realization calls for bright and high-purity high-$d$ entangled photon sources, transmission channels with low loss and minimal crosstalk, and measurement systems capable of stable joint detections. Despite these challenges, the security motivation remains strong~\cite{erhard2018twisted}. By the monogamy of entanglement~\cite{coffman2000distributed}, if Alice and Bob share a nearly maximally entangled state, Eve is effectively decoupled from their correlations. Consequently, the observed reduction from maximal entanglement, or equivalently the strength of a Bell-inequality violation, provides a direct upper bound on Eve’s information and determines the required privacy-amplification cost.

{
\color{black}
\subsection{Common Experimental Benchmarks and Representative Setups}\label{sec3_add}

The experimental development of OAM-encoded QKD has progressed from proof-of-principle laboratory demonstrations to more realistic implementations, including in-field experiments and technological prototypes. This evolution can be systematically interpreted using two complementary dimensions \cite{pirandola2020advances}: experimental environment and system completeness. The experimental environment characterizes the channel conditions under which a system is evaluated, ranging from controlled laboratory settings to practical scenarios such as fiber links or turbulence-affected free-space channels. System completeness reflects the extent to which the full QKD workflow is implemented, thereby providing an indicator of technological maturity.

To complement the protocol-based classification in the preceding subsections, Tab.~\ref{tab:oam_hdqkd_comparison} introduces a benchmark-oriented framework for comparing representative OAM-encoded HD-QKD demonstrations. Unlike Tabs.~\ref{pam_qkd} and~\ref{en_qkd}, which organize studies by protocol type, this benchmark emphasizes experimental environment, system completeness, representative setups, SKR interpretation, and primary evaluation focus, thereby enabling a more systematic comparison across different implementations.

As discussed in Section~\ref{sub_QKD}, the QKD workflow includes qubit exchange, sifting, parameter estimation, error correction, and privacy amplification. For implementation-oriented comparison, we adopt a broad practical notion of the physical layer to refer to experiments that realize state preparation, transmission, detection, and basic statistical characterization up to parameter estimation, without necessarily achieving a full end-to-end QKD protocol.

To the best of our knowledge, OAM-encoded QKD does not yet have a standardized dataset or a universally accepted benchmark suite. Consequently, cross-study comparison relies primarily on consistent reporting of experimental conditions, system architectures, measured physical quantities, and SKR interpretation. From this perspective, the current experimental landscape reveals a meaningful but uneven progression: while many demonstrations remain at the proof-of-principle stage in terms of system completeness, the experimental environment has steadily expanded toward more realistic scenarios, including in-field fiber links and turbulence-affected free-space conditions \cite{groblacher2006experimental, zhao2013large, mirhosseini2015high, larocque2017generalized, chang2023compact, sun2024experimental, wang2019characterizing, cozzolino2019orbital, scarfe2025fast}.

}
\setlength{\cmidrulewidth}{0.02em}
\begin{table*}[!t]
\captionsetup{font={color=black}}
\color{black}
\centering
\scriptsize
\setlength{\tabcolsep}{3pt}
\renewcommand{\arraystretch}{1.0}
\caption{Comparative summary of representative OAM-encoded HD-QKD demonstrations.}
\label{tab:oam_hdqkd_comparison}
\newcolumntype{M}[1]{>{\centering\arraybackslash}m{#1}}
\newcolumntype{L}[1]{>{\raggedright\arraybackslash}m{#1}}
\resizebox{\textwidth}{!}{
\begin{tabular}{L{1.5cm} L{2.0cm} L{2.0cm} L{3.0cm} L{3.0cm} L{2.2cm}}
\toprule
\textbf{Reference} & \textbf{Experimental environment} & \textbf{System completeness} & \textbf{Representative setup} & \textbf{SKR interpretation} & \textbf{Primary evaluation focus} \\
\midrule
\cite{groblacher2006experimental} 
& Controlled laboratory
& Only physical layer
& Spontaneous parametric down-conversion (SPDC) entanglement source with hologram-based projective detection.
& No SKR reported
& Entanglement-based security verification. \\
\cmidrule(lr){1-6}
\cite{zhao2013large}
& Controlled laboratory 
& Only physical layer 
& SPDC entanglement source with SLM-based sector-state analysis and coincidence detection.
& No SKR reported
& Large-alphabet discrimination. \\
\cmidrule(lr){1-6}
\cite{mirhosseini2015high}
& Controlled laboratory
& Only physical layer
& Digital micromirror device (DMD)-based OAM encoding with a log-polar mode-sorter receiver.
& Theoretical SKR inferred from measured error statistics using the formula from \cite{cerf2002security}.
& Implementation feasibility of OAM-encoded QKD. \\
\cmidrule(lr){1-6}
\cite{larocque2017generalized}
& Controlled laboratory
& Only physical layer
& Spatial light modulator (SLM)-based OAM encoding with PBOE-assisted OAM sorting and camera detection. 
& Theoretical SKR inferred from measured error statistics using the formula from \cite{shor2000simple}.
& Feasibility assessment for PBOE-based sorting. \\
\cmidrule(lr){1-6}
\cite{chang2023compact}
& Controlled laboratory
& Only physical layer  
& SLM-based OAM encoding with an integrated mode-converter receiver. 
& Theoretical SKR inferred from measured error statistics using the formula from \cite{chang2022security}.
& MPUB-based security analysis. \\
\cmidrule(lr){1-6}
\cite{sun2024experimental}
& Controlled laboratory 
& Only physical layer
& Spiral-phase-plate for OAM encoding with grating-assisted OAM detection.
& Theoretical SKR inferred from measured error statistics using the formula from \cite{lo2005decoy}.
& Key-rate-oriented evaluation. \\
\cmidrule(lr){1-6}
\cite{wang2019characterizing}
& Controlled laboratory
& Only physical layer 
& q-plate-based spin-OAM hybrid encoding and decoding.
& Theoretical SKR inferred from measured error statistics using the formula from \cite{lo2005decoy}.
& Implementation feasibility of OAM-encoded HD-QKD. \\ 
\cmidrule(lr){1-6}
\cite{cozzolino2019orbital}
& In-field 
& Only physical layer
& Vortex-plate-based spin--OAM hybrid encoding over air-core fiber.
& Theoretical SKR inferred from measured error statistics using the formula from \cite{islam2017provably}.
& Implementation feasibility of fiber transmission for OAM-encoded photons. \\
\cmidrule(lr){1-6}
\cite{scarfe2025fast}
& Laboratory emulation of realistic turbulence 
& Only physical layer
& SLM-based OAM encoding with a copropagating reference beam and AO-assisted free-space detection.
& Theoretical SKR inferred from measured error statistics using the formula from \cite{bouchard2018experimental}. 
& AO-assisted free-space OAM-encoded HD-QKD transmission feasibility. \\
\bottomrule
\end{tabular}}
\end{table*}

\section{System-Level Challenges in OAM-Encoded HD-QKD}\label{challenges} 
\begin{figure}
    \centering
    \includegraphics[width=1\linewidth]{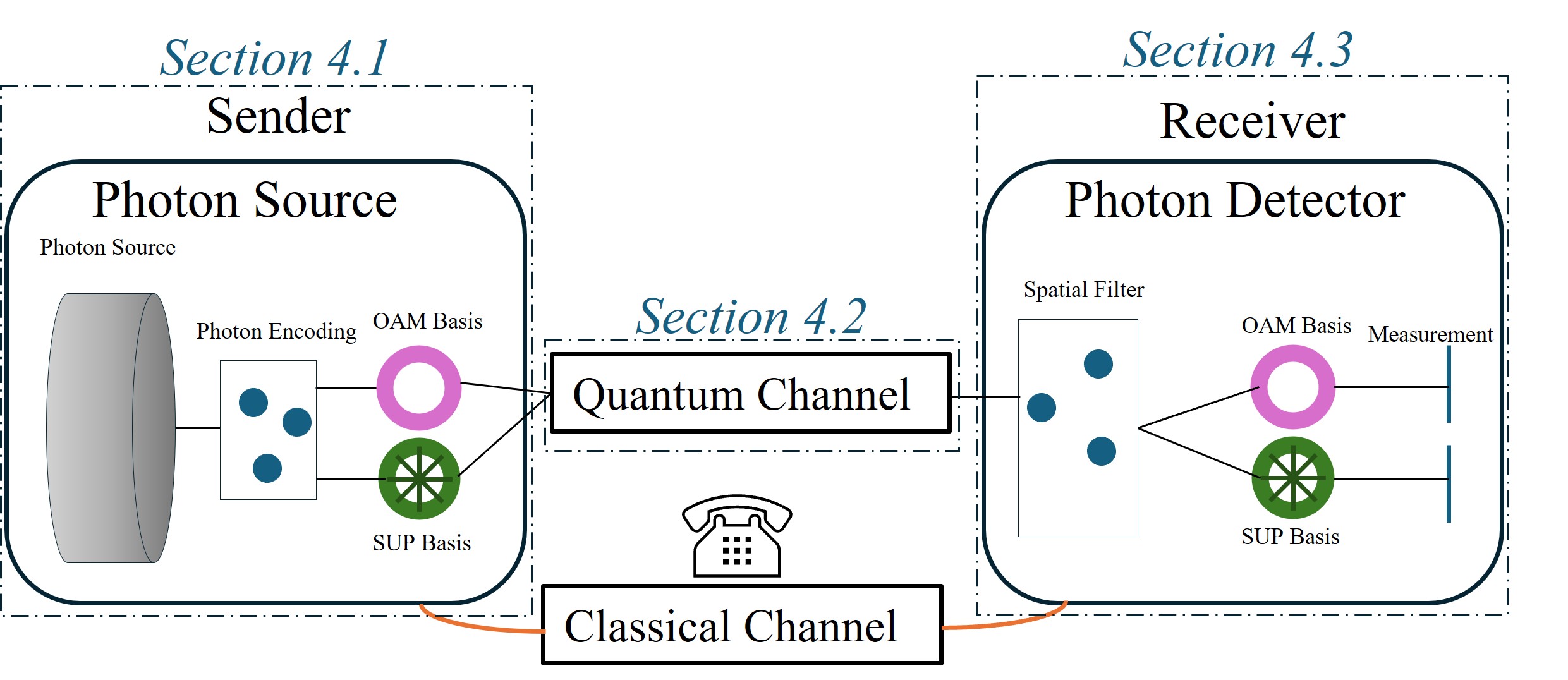}
    \caption{System-stage overview of the challenges in OAM-encoded QKD.}
    \label{fig:challenge}
\end{figure}

Fig.~\ref{fig:challenge} provides a system-level view of an OAM-encoded QKD link, outlining the three stages that structure this section: the sender, the quantum channel, and the receiver. These stages correspond to Sections~\ref{Generation}, \ref{Transmission}, and \ref{Reception}, respectively. The sender stage is scheme-dependent. In prepare–measure protocols, the sender is Alice, who prepares and transmits the quantum states. In 
entanglement-based protocols, the sender is the entangled-photon source, which distributes one photon to Alice and the other to Bob. Similarly, the receiver stage is Bob in prepare–measure systems, whereas in entanglement-based settings it refers to the detection modules used by both Alice and Bob.

The remainder of this section follows this system-stage organization. Section~\ref{Generation} examines challenges that arise during state generation, including the preparation of high-dimensional OAM modes and their superpositions.  
Section~\ref{Transmission} then analyzes impairments introduced by the quantum channel and methods to mitigate them. Finally, Section~\ref{Reception} considers imperfections at the detection stage and discusses strategies for compensating these effects.

\subsection{Challenges in State Generation}\label{Generation}
On the sender side, OAM-encoded QKD faces several fundamental and practical challenges associated with preparing high-dimensional spatial states. Generating OAM modes and their superpositions with high efficiency, stability, and rapid switching is technically demanding, and these requirements become more stringent as the dimensionality of the encoding grows. 
For clarity, we group the main challenges in state generation into four categories:
(i) sensitivity to reference-frame misalignment, (ii) the trade-off between dimensionality and SKR, 
(iii) hardware limitations in mode preparation, and (iv) the practical limits of scaling to very high 
dimensions.

\subsubsection{Sensitivity to Reference-Frame Misalignment}

A prominent difficulty in free-space QKD is the sensitivity of OAM modes to reference-frame misalignment. Because OAM photons carry a helical phase structure, any relative rotation between the transmitter and the receiver alters the observed mode and can scramble the encoding. Maintaining a stable shared reference orientation therefore introduces 
significant operational overhead~\cite{d2012complete}.  

A widely used solution is hybrid encoding, in which multiple degrees of freedom are combined to produce rotation-invariant states. For example, photons can be encoded jointly in polarization and OAM using devices such as q-plates, enabling states that remain unchanged under rotations around the propagation axis~\cite{vallone2014free}. These 
polarization-OAM hybrid states significantly reduce the impact of misalignment and other spatial perturbations. Alignment-free QKD without active orientation control—has been demonstrated experimentally, including over metropolitan-scale free-space links~\cite{d2012complete, vallone2014free}.

\subsubsection{Dimensionality, Per-Photon Capacity, and SKR Trade-Offs}

Although HD-QKD can associate more bits with each signal photon, increasing the dimensionality of the encoding does \emph{not} automatically improve the SKR. The relevant performance indicator is the SKR, defined as the number of secure bits generated per unit time. Even if a single photon carries more secure information, the throughput of the 
system—how many photons can be generated, transmitted, and detected per second—ultimately determines the achievable SKR.

Fundamental capacity limits constrain how much secure information a single photon can carry, regardless of how many spatial or temporal modes are packed into its structure~\cite{xu2020secure}. As a result, using many modes to form a large symbol may increase the per-photon key rate but cannot surpass the per-photon capacity bound. In many cases, transmitting the same modes as separate photons and detecting them in parallel with sufficiently fast hardware yields a higher aggregate SKR. Consequently, high-dimensional OAM encodings can appear advantageous on a per-photon basis, yet fail to deliver a net benefit once detector speeds, optical loss, and clock-rate constraints are accounted for. A central challenge is therefore to balance dimensionality with practical throughput to ensure a genuine improvement in SKR \cite{ogrodnik2025high}.

\subsubsection{Hardware Limitations in OAM State Preparation}

The physical generation of OAM modes is also limited by current optical hardware. Conventional approaches such as spatial light modulators (SLMs) and diffractive holograms often introduce substantial optical loss and provide limited switching speed. An SLM, for example, diffracts only part of the incident field into the target OAM mode, with the 
remaining power distributed into unwanted diffraction orders or absorbed, resulting in transmitter insertion loss. Because each decibel of loss directly reduces the SKR, such losses are particularly problematic for quantum communication.

Speed is another major constraint. Symbol-by-symbol OAM switching with SLMs typically requires between $1$\,ms and $1$\,\textmu s per pattern update—far slower than the nanosecond-scale polarization modulation used in many standard QKD systems. This slow reconfiguration prevents random switching among large OAM alphabets at high clock rates.

Recent developments aim to overcome these limitations. Electrically tunable q-plates can map polarization states to OAM with high efficiency and microsecond-scale switching speeds \cite{d2012complete}, enabling stable generation of polarization--OAM hybrid states without requiring interferometric stability. Integrated photonic circuits~\cite{butow2024generating} and chip-based mode generators~\cite{zhao2025all} offer further improvements, including lower loss, enhanced stability, smaller footprints, and fast electrically driven reconfiguration. These advances represent important steps toward practical, high-rate OAM transmitters.

\subsubsection{Practical Limits on Scaling to Very High Dimensions}

Although the OAM degree of freedom spans an unbounded Hilbert space in principle, 
experimental control becomes increasingly difficult as the mode index $l$ grows.  
Generating and distinguishing high-$l$ modes with high fidelity is challenging: small 
imperfections in state preparation—such as SLM misalignment~\cite{wang2021high} or residual 
optical aberrations—introduce mode crosstalk. Detection systems face similar limitations, 
since mode sorters must resolve increasingly fine spatial features as $l$ increases. 
Existing devices therefore impose an effective upper bound on the dimensionality that can 
be coherently controlled in OAM-based QKD~\cite{cozzolino2019high}.

Most experiments to date have operated at moderate dimensions (e.g., $d = 3, 4, 8$), where 
error rates remain manageable. Pushing to much higher-dimensional alphabets requires 
proportionally more calibration effort and is susceptible to diminishing returns: error 
accumulation often offsets the nominal information gain~\cite{erhard2018twisted, 
cozzolino2019high}. As a result, the state-generation stage of OAM-encoded QKD must operate 
within practical limits on the usable dimension. Increasing dimensionality is beneficial only 
up to a point, beyond which noise, optical complexity, and stability constraints outweigh the 
potential advantages~\cite{zhao2025all}. Determining this operating point—and improving the 
underlying devices so that it can be pushed higher—remains an active area of research.

\subsection{Challenges in Quantum Transmission}\label{Transmission}
While OAM encoding provides access to a large state space for QKD, it also introduces 
significant difficulties during transmission. OAM-encoded photons are highly susceptible to 
distortions imposed by the quantum channel, and these impairments differ substantially between 
free-space and fiber links. To organize the discussion, we group the dominant challenges into 
three categories: (i) mode crosstalk, (ii) engineered-fiber dispersion effects, and (iii) mode-dependent 
diffraction.

\subsubsection{Mode Crosstalk in Free-Space and Fiber Links}

In practice, crosstalk is one of the most severe impairments affecting OAM-encoded QKD. In 
this paper, crosstalk refers to the situation in which the OAM mode prepared by Alice is 
detected as a different mode at Bob, resulting in symbol misidentification at the receiver.

\textbf{Free-space crosstalk:} 
In free-space links, atmospheric turbulence and pointing jitter distort the optical 
wavefront of OAM photons, causing unwanted coupling between modes and inducing beam wander at 
the receiver~\cite{erhard2018twisted}. A field test conducted in 2017 over a 300\,m 
intra-city link using ququarts ($d=4$) observed pronounced mode crosstalk and increasing beam 
divergence~\cite{sit2017high}. Instead of deploying AOs, the experiment used an 
idler photon—called a "weather-vane"—as a coaxial turbulence reference. Frames with 
abnormally low idler counts were discarded during post-processing, implementing a 
turbulence-gated detection strategy. This post-selection reduced the QBER to roughly 
$11\%$ (below the $18.9\%$ threshold for $d=4$) and produced an information yield of 
$0.65$\,bit/sifted photon, although at the cost of reduced throughput since many frames were 
discarded. Extending this approach to longer, more turbulent links would likely require 
AOs. Despite progress, multi-kilometer OAM-encoded HD-QKD in free space remains elusive 
because turbulence and pointing jitter rapidly degrade mode purity. Although OAM 
correlations have been observed over much longer distances (e.g., about $143$\,km 
in~\cite{krenn2016twisted}), the level of crosstalk and attenuation still prevents their use 
for HD-QKD~\cite{erhard2018twisted}.

\textbf{Fiber inter-modal crosstalk:}  
Optical fibers introduce a fundamentally different type of impairment. Standard telecom 
single-mode fiber cannot support OAM modes at all, so most QKD demonstrations have relied on 
other degrees of freedom. Multi-mode and few-mode fibers can, in principle, carry OAM modes, but even small
perturbations, such as bending, stress, or index fluctuations, induce strong
inter-modal coupling and rapidly scramble the transmitted OAM information~\cite{cozzolino2019orbital,wang2021high}.
 Even a few meters of uncontrolled multi-mode fiber can generate substantial 
mode crosstalk. A study conducted in 2017~\cite{ding2017high} emphasized that although OAM-encoded QKD 
had been demonstrated in free-space settings, long-distance fiber transmission of OAM modes 
remained extremely challenging due to this inter-modal crosstalk.

\subsubsection{Engineered Fibers and Mode-Dependent Dispersion}

To mitigate inter-modal coupling, researchers have explored specially engineered fiber 
designs that preserve OAM mode structure. Air-core (hollow-core) fibers and ring-core fibers 
are designed so that pairs of OAM modes with opposite spin–orbit orientation experience 
slightly different effective refractive indices~\cite{cozzolino2019orbital, wang2021high}. This 
spectral separation helps maintain orthogonality and suppresses unwanted coupling over longer 
distances.

Even in these fibers, however, different OAM modes propagate with different group velocities 
because their spatial field profiles differ. This mode-dependent dispersion causes pulsed OAM 
signals to arrive at different times, which degrades interference visibility unless 
compensated. A 1.2\,km experiment~\cite{cozzolino2019orbital} required optical delay lines to 
realign the arrival times of $l=6$ and $l=7$ components. Furthermore, hollow-core fibers still 
exhibit finite attenuation (several dB/km), limiting both the transmission distance and the 
achievable SKR—particularly important for HD-QKD, which has tighter loss requirements.

Although continued advances include custom OAM-preserving fibers and multicore fibers that 
isolate spatial channels, fiber-based OAM-encoded QKD currently requires delicate and costly 
infrastructure and remains restricted to relatively short spans. These limitations have 
motivated exploration of alternative spatial encodings (for example, multicore path-encoded 
QKD), which avoid the fragility of OAM modes in fibers and have achieved HD-QKD over tens of 
kilometers.

\subsubsection{Mode-Dependent Diffraction and Same-Order Encoding}

A third transmission challenge arises from mode-dependent diffraction, also referred to as 
state-dependent diffraction~\cite{zhao2019performance, li2024state}. During free-space 
propagation, OAM modes diverge at rates that increase with the mode index $l$: higher-order 
modes spread more rapidly and develop broader intensity profiles. When multiple OAM modes or 
superpositions are transmitted, each state experiences a different amount of beam spread and 
loss, complicating long-distance propagation. A related effect appears in fibers, where 
different transverse mode profiles lead to mode-dependent confinement and attenuation.

To address this issue, Wang \emph{et al.}~\cite{wang2020high} proposed an HD-QKD protocol 
based on mutually partially unbiased bases (MPUBs). By encoding information in same-order 
spatial modes, which share similar diffraction behavior, the protocol avoids 
mode-dependent broadening and increases robustness during propagation. A 2023 proof-of-principle 
experiment~\cite{chang2023compact} confirmed the feasibility of distributing high-dimensional 
OAM states in $2$-, $4$-, $5$-, and $6$-dimensional Hilbert spaces using only linear-optical 
components, marking an important step toward practical free-space HD-QKD.

\subsubsection{Summary of Transmission Challenges}

In summary, achieving robust and scalable transmission of OAM-encoded quantum keys remains 
a significant challenge. Free-space links are primarily limited by atmospheric turbulence and 
pointing instability, while fiber links are constrained by inter-modal coupling and 
mode-dependent dispersion. Both settings must also contend with differential losses and other 
implementation overhead. Active research efforts—including AO methods, custom 
fiber designs, improved mode-sorting technologies, and high-dimensional error-correction 
strategies—continue to advance the field. With sustained progress, OAM and other structured 
photonic modes may eventually support quantum communication over metropolitan and, 
ultimately, global scales~\cite{erhard2018twisted}.

\subsection{Challenges in Detection and Measurement}\label{Reception}

The receiver plays a critical role in OAM-encoded QKD, especially in free-space links where 
turbulence, pointing errors, and diffraction collectively degrade detection efficiency and 
induce mode crosstalk. A recent receiver-side study~\cite{li2024state} showed that 
state-dependent diffraction, turbulence-induced mode distortions, and mode mismatch at the 
aperture can all reduce the fidelity with which the intended OAM symbol is recovered. 
Consequently, significant research has focused on receiver-side compensation methods, which 
can be grouped into four main approaches: (i) co-design of the modulation alphabet and 
measurement bases, (ii) AO-based correction of wavefront distortions, (iii) machine-learning-based inference of turbulent phase screens and \textcolor{black}{(iv) nonlinear-optical phase-conjugation-based compensation. }

\subsubsection{Co-Design of Modulation Alphabets and Measurement Bases}
When receiver imperfections such as mode mismatch, finite-aperture loss, and 
basis-dependent phase errors dominate, performance can be improved by designing 
the information alphabet and the measurement bases to share common propagation behavior. One representative 
strategy~\cite{chang2023compact} replaces fully MUBs with mutually 
partially unbiased bases (MPUBs) constructed from same-order LG modes and 
$45^\circ$-rotated Hermite–Gaussian (HG) modes. All symbols obey a common mode order 
$N=2p+|l|=n+m$, giving a Hilbert-space dimension $d=N+1$.  

By construction, these same-order modes share identical divergence and accumulate equal 
Gouy-phase shifts during propagation. This suppresses mode-dependent loss and the 
basis-dependent phase walk-off that otherwise arises between OAM and SUP (superposition) 
bases over finite apertures. On the receiver side, a $\pi/2$ mode converter maps HG states 
into LG states, so a single LG sorter suffices for both bases. Basis selection is performed 
with a polarization-controlled active gate, reducing the required number of detectors by half. 
Proof-of-principle demonstrations have validated operation for $d = 2, 4, 5,$ and $6$, 
together with an experimental analysis of basis-dependent detection imperfections 
\cite{chang2023compact}.

A complementary alphabet-design approach~\cite{zhu2021compensation} targets invariants that 
are resilient to turbulence. Using the spatial–polarization differential phase-shift keying 
(SPDPSK) method, vector–vortex states are formed from superpositions of opposite-charge LG 
modes in the two circular polarizations. The decoding optics use anisotropic phase plates that 
impose order-dependent phase shifts $\Delta\phi_n = 2n\theta$ followed by balanced detection. 
This produces a decision metric near $\pm 1$ for the correct OAM order and near $0$ otherwise. 
Because turbulence is largely polarization-insensitive, these vector–vortex states maintain 
their spatial–polarization structure even when the full complex field is distorted. Using SPDPSK, compensation-free transmission in a $34$-dimensional alphabet 
(corresponding to an information payload of $5.09$ bit/photon) was demonstrated at 
scintillation indices up to $1.54$ (a standard measure of turbulence strength), 
highlighting the advantages of embedding propagation physics directly into the symbol alphabet.

\subsubsection{AOs for Real-Time Wavefront Correction}

AO provides a physical method for real-time compensation of turbulence. 
AO has long been used to correct atmospheric distortions in astronomical imaging 
\cite{van2004performance}, but its application to free-space quantum communication is more 
recent. In the context of OAM-encoded QKD, AO systems aim to recover the intended spatial mode 
by dynamically flattening or reshaping the distorted wavefront.

Most prior studies investigated AO for QKD at a theoretical level 
\cite{leonhard2018protecting, sorelli2019entanglement}. Experimental progress, however, has 
been encouraging. Scarfe \emph{et al.}~\cite{scarfe2025fast} demonstrated that high-speed AO 
driven by a 97-actuator deformable mirror significantly reduced mode crosstalk and lowered 
the error rate in a turbulent free-space HD-QKD link. This work shows that with sufficient 
correction bandwidth and spatial resolution, AO can convert otherwise insecure channels into 
secure ones.

\subsubsection{Machine-Learning-Based Compensation of Turbulence}

A third receiver-side strategy uses machine learning (ML) to infer and cancel turbulence-induced wavefront aberrations. Liu \emph{et al.}~\cite{liu2019deep} developed a convolutional neural network (CNN) that takes paired intensity images of a co-propagating Gaussian probe beam (with and without turbulence) and estimates an equivalent turbulent phase screen. The receiver then loads the inverse phase pattern onto an SLM to correct the simultaneously transmitted OAM channels.

Because the CNN learns a direct mapping from probe intensity to phase without iterative wavefront reconstruction, the method operates quickly enough for dynamic links and generalizes across different turbulence strengths and propagation distances. In system-level link testing, machine-learning-based compensation restored clean OAM intensity profiles, suppressed mode crosstalk, and produced tighter constellation diagrams with substantially reduced QBER. These results indicate that deep learning can serve as a compact digital AO module for OAM-encoded free-space quantum links.

\subsubsection{\textcolor{black}{Nonlinear-Optical Phase-Conjugation-Based Compensation}}

\textcolor{black}{A fourth line of research investigates nonlinear-optical phase conjugation for turbulence compensation. In contrast to co-design approaches, which enhance robustness by selecting symbols and measurement bases with similar propagation characteristics, this approach does not rely on robustness-by-design. It also differs from AO and machine-learning-based methods, which perform turbulence mitigation through channel estimation or inference followed by externally applied wavefront correction \cite{van2004performance,scarfe2025fast}. Instead, compensation is generated intrinsically within the optical interaction itself \cite{santos2025experimental}.}

\textcolor{black}{A representative implementation is based on stimulated parametric down-conversion (StimPDC). In this scheme, a distorted probe beam seeds a $\chi^{(2)}$ nonlinear process, and the generated idler field carries both the intended spatial mode and the phase-conjugated information of the turbulent channel \cite{santos2025experimental}. Consequently, turbulence compensation is embedded directly in the field-generation process, distinguishing this approach from both robustness-by-design strategies and conventional estimate-and-correct methods.}

\textcolor{black}{Aguilar \emph{et al.} \cite{aguilar2026all} proposed an all-optical StimPDC-based compensation scheme for HD-QKD with transverse spatial modes. Although the framework was presented for general spatial-mode encoding, the demonstrated MUB states were constructed from superpositions of LG modes, thereby explicitly incorporating OAM components.}

\section{Emerging Directions for OAM-Encoded HD-QKD}\label{future}

QKD has advanced rapidly in recent years, and the unique properties of OAM have motivated 
its integration into several emerging QKD architectures. These include TF-QKD, which is designed to extend the achievable communication distance~\cite{wang2022twin}; CV-QKD, which aims to increase aggregate key rates through high-rate optical detection~\cite{fan2022robust}; and MDI-QKD or DI-QKD, which enhances implementation security by removing trust assumptions on the 
detectors~\cite{jo2025security}. Together, these developments reflect QKD’s core objectives 
\cite{pirandola2020advances}: achieving longer distances, higher secure throughput, and 
stronger security guarantees.

\subsection{TF-QKD with OAM}\label{sub_TF}

QKD provides a theoretically proven method for distributing secret keys, yet channel loss 
remains the principal limitation for long-distance operation because quantum signals cannot 
be amplified. For a pure-loss channel with transmittance $\eta$, the parameter $\eta$
represents the fraction of optical power that survives transmission.
The corresponding repeaterless secret key capacity per channel use
(equivalently, per optical mode) is given by the Pirandola--Laurenza--
Ottaviani--Bianchi (PLOB) bound~\cite{pirandola2017fundamental}, which
scales as $1.44\,\eta$ for $\eta \ll 1$. This fundamental limit applies to all repeaterless QKD protocols and 
encodings, including discrete- and continuous-variable schemes, qubit and high-dimensional 
encodings, and widely deployed protocols such as decoy-state BB84 and MDI-QKD. Although 
MDI-QKD removes detector-side vulnerabilities by placing the measurement device at an 
untrusted relay, it still requires well-characterized, trusted transmitters. In the absence 
of practical quantum repeaters, long-haul deployment often relies on multiple trusted 
relays, which has been demonstrated in field trials but increases both the operational cost 
and the system’s security exposure as relays proliferate.

TF-QKD was introduced to address this constraint and enable practical 
long-distance quantum communication without quantum memories~\cite{lucamarini2018overcoming}. 
Unlike conventional point-to-point schemes, TF-QKD achieves an 
$\mathcal{O}(\sqrt{\eta})$ scaling of the SKR, allowing significantly 
longer transmission distances and reducing the required number of trusted nodes. Field and 
laboratory demonstrations of the sending-or-not-sending (SNS) variant have confirmed these 
performance gains: with full finite-size analysis, secure key exchange has been achieved 
beyond $500\,\mathrm{km}$ and up to $658\,\mathrm{km}$ in ultra-low-loss fiber, with other 
implementations reporting positive rates around $615.6\,\mathrm{km}$~\cite{lucamarini2018overcoming}. 
These results validate the predicted repeater-like behavior of TF-QKD while remaining 
consistent with the PLOB limit on a per-mode basis.

Conventional TF-QKD encodes information in the optical phase, which requires Alice and Bob 
to maintain tight phase alignment; even slow reference-frame drift translates directly into 
errors. To remove this fragility, Meng \emph{et al.}~\cite{meng2020twin} proposed replacing 
the optical phase with the rotation-invariant OAM degree of freedom, yielding a 
reference-frame-independent (RFI) TF-QKD protocol. Because OAM modes remain unchanged under 
rotations around the propagation axis, the protocol becomes insensitive to relative 
orientation between the parties’ frames. This removes the need for active phase stabilization 
and eliminates intrinsic misalignment errors—an important advantage for free-space TF-QKD 
links where atmospheric turbulence dominates over slow phase drift.

Within the broader TF-QKD family, phase-matching QKD (PM-QKD) reformulates central 
interference in a preparation--measurement framework. Alice and Bob independently prepare 
weak coherent states, randomize their phases, and transmit them to an untrusted relay 
(Charlie); raw key bits are kept only from detection events satisfying the phase-matching 
condition $\phi_A \approx \phi_B$. In this way, PM-QKD maintains the favorable 
distance scaling of TF-QKD while inheriting the measurement-device-independence of MDI-QKD. 
Operationally, the core steps are independent phase randomization, interference at Charlie, 
and sifting on phase-matched events.

Shen \emph{et al.}~\cite{shen2022phase} implemented an OAM-based PM-QKD variant in which the 
logical bit value is encoded in the relative phase of a coherent-state superposition of two 
OAM modes with opposite topological charges. Successful events at Charlie correspond to 
relative phases of $\pm \pi/2$, after which phase post-selection yields the raw key. The 
security was analyzed using an equivalent entanglement-distillation model, and performance 
was evaluated over turbulent free-space channels. Numerically, the authors reported that the key-rate bound of the original PM-QKD protocol can be surpassed at distances beyond approximately $230\,\mathrm{km}$ 
(overall loss exceeding $45\,\mathrm{dB}$) and that the maximum secure distance exceeds that 
of the original PM-QKD protocol, largely due to the absence of basis misalignment. Taken 
together with earlier demonstrations of reference-frame-independent TF-QKD using OAM, these 
results show that replacing a fragile optical phase reference with a rotation-invariant OAM 
encoding preserves central interference, supports an untrusted measurement node, and offers 
a practical benefit for free-space twin-field links subject to turbulence.

\subsection{CV-QKD with OAM Multiplexing}\label{sub_CV}

CV-QKD using coherent states~\cite{grosshans2002continuous} has 
experienced renewed interest due to its natural compatibility with existing telecommunications 
infrastructure, including commercial continuous-wave lasers and coherent receivers. In general, 
QKD protocols can be grouped into discrete-variable (DV) and continuous-variable (CV) classes. 
Unlike DV or qubit-based approaches, CV-QKD encodes secret keys in the quadratures of the 
electromagnetic field and recovers them using coherent detection. Coherent receivers are 
attractive for practical deployment because they operate at room temperature, offer high 
detection efficiencies, and align well with standard telecom hardware.

In DV-QKD with OAM, a finite set of OAM indices forms a high-dimensional alphabet, and single 
photons are measured using projective detection in OAM and SUP bases. In contrast, OAM-based 
CV-QKD treats distinct OAM modes as parallel CV channels. Each mode carries independent, 
Gaussian-modulated quadratures that are read out using mode-matched homodyne or heterodyne 
detection. This approach enables spatial-mode multiplexing, creating the possibility of 
significant aggregate-rate scaling. At the same time, it introduces challenges specific to 
CV protocols.

Maintaining mode fidelity in realistic channels is a primary difficulty for OAM-based 
CV-QKD. In free-space links, atmospheric turbulence distorts the spatial wavefront, causing 
mode crosstalk and phase noise~\cite{ruan2021high}. Ruan \emph{et al.}~\cite{ruan2021high} 
developed an end-to-end model incorporating turbulence-induced crosstalk, effective 
transmittance, excess noise, and SKR, and used it to evaluate discrete-modulated CV-QKD with 
OAM multiplexing over turbulent free-space channels. Their simulations showed that OAM multiplexing can boost the total SKR, with theoretical 
peak rates near $38.31\,\mathrm{Mbit/s}$, although the maximum distance remains limited 
by turbulence and by the use of higher-order OAM modes.
 Higher-order OAM modes are especially vulnerable because even moderate 
turbulence blurs the characteristic ring-shaped intensity profiles of LG 
modes and distorts their helical phase structure. As a result, OAM-based CV-QKD is best suited 
to short-range, high-throughput free-space links.

Noise sensitivity presents another critical issue. CV-QKD requires low-noise channels to 
retain security, yet experiments indicate that OAM-bearing continuous-variable states can lose 
quantum correlations abruptly under excess noise. Zeng \emph{et al.}~\cite{zeng2022deterministic} 
generated OAM-multiplexed CV entanglement and studied its behavior under controlled loss and 
noise. In pure-loss channels, OAM entanglement persisted comparably to the Gaussian 
($l=0$) mode, indicating that attenuation alone does not significantly penalize OAM. 
However, with modest excess noise the situation changed: higher-order modes lost entanglement 
suddenly, with $l=2$ exhibiting a sudden death of entanglement and steering. This highlights 
the strong sensitivity of OAM modes to phase noise and mode-dependent fluctuations, even when 
loss remains tolerable. A related study found that some coherence measures can survive even 
after entanglement vanishes~\cite{wen2022quantum}, but for QKD any increase in excess noise 
reduces the secrecy rate. Thus, stabilizing OAM modes against phase perturbations, for 
example via real-time wavefront correction or noise-whitening techniques, is essential for 
keeping the overall noise budget within secure bounds.

Beyond the transmission channel, the practical generation and detection of OAM-encoded CV 
states pose additional challenges. On the transmitter side, preparing squeezed or entangled 
OAM modes with high purity is non-trivial. Early demonstrations required careful optical 
engineering. For example, Liu \emph{et al.}~\cite{liu2014experimental} produced 
OAM-bearing CV hyperentanglement using a specially tuned optical parametric oscillator that 
simultaneously squeezed two polarization and two OAM quadratures; compensating astigmatism in 
the cavity was necessary to obtain clean mode pairing. More recent proposals have explored alternatives to optical parametric oscillators, including 
atomic-vapor four-wave mixing processes.
 Meng \emph{et al.}~\cite{meng2025continuous} 
proposed a scheme for generating OAM-entangled CV beams through Raman processes in a 
coherently prepared medium, offering a possible route toward bright multi-OAM entangled 
sources, although such approaches remain at an early stage.

On the receiver side, detecting OAM-encoded quadratures requires advanced optical processing. 
A standard homodyne receiver must mode-match its local oscillator to each incoming OAM mode. 
Parallel detection across multiple OAM channels may require dynamic holograms or mode 
converters to transform each OAM mode into a Gaussian mode prior to detection; any 
imperfection in conversion or alignment manifests as additional noise or loss. In a 
high-speed OAM-multiplexed CV-QKD experiment, Qu \emph{et al.}~\cite{qu2017high} employed a 
phase-noise cancellation module to stabilize the relative phase in each OAM channel and 
maintain local-oscillator lock. Even with stabilization, the authors reported that each OAM 
mode required a minimum link transmission (for instance, above $21\%$ for $l=2$) to remain 
secure; below this threshold, excess noise overwhelmed the quantum signal. This illustrates 
the narrow operational margins when using complex spatial modes for CV-QKD.

In summary, OAM-encoded CV-QKD offers the promise of high-dimensional encoding and potentially very high aggregate SKRs, but several practical hurdles remain. Atmospheric turbulence and mode crosstalk limit transmission distance, higher-order OAM modes are particularly sensitive to excess noise, and both state generation and detection require sophisticated mode control. Ongoing research explores AOs and machine-learning-based control to stabilize OAM modes in real time, as well as multiplexing strategies that favor the most robust modes. As experimental demonstrations of entangled OAM links mature in realistic atmospheric conditions, the trade-offs between dimensional gain, noise tolerance, and system complexity will become clearer, guiding the development of stable and high-rate OAM-based CV-QKD systems.

\subsection{OAM-Enabled MDI/DI-QKD}\label{MDI}

Although QKD is information-theoretically secure, practical implementations 
inevitably deviate from the ideal conditions assumed in security proofs. Imperfections in 
sources, modulators, and especially detectors open side channels that can be exploited by 
quantum hacking attacks~\cite{rubenok2013real}. Two leading approaches that directly address 
these implementation loopholes are MDI-QKD and DI-QKD.

DI-QKD represents the strongest form of implementation security. In DI-QKD, Alice and Bob do 
not need to trust the internal operation of their devices; instead, security is certified by 
observing a Bell-inequality violation using entangled states~\cite{zhang2022device}. This 
ensures that any deviation from the ideal behavior is detected through the Bell test itself. 
However, DI-QKD is extremely demanding experimentally because it requires near-lossless 
detection and high-quality entanglement distribution. To date, device-independent key 
exchange has been demonstrated only under tightly controlled laboratory conditions, and 
scaling such demonstrations to practical link distances remains an open challenge.

MDI-QKD provides a more experimentally accessible route to closing detector-side 
vulnerabilities. In MDI-QKD, Alice and Bob send quantum states to an untrusted relay 
(Charlie), who performs a Bell-state measurement. Since all detection happens at an 
untrusted node, the users’ devices become immune to detector side-channel attacks, and 
secure key exchange can still be achieved even if Charlie is fully adversarial~\cite{sekga2023measurement}. 
A wide range of MDI protocols have been proposed—including twin-field variants for 
long-distance links—with encodings in polarization~\cite{ferreira2013proof}, phase~\cite{tamaki2012phase}, 
and time-bin degrees of freedom~\cite{nie2016experimental}. In all cases, the essential 
feature is that the measurement unit need not be trusted.

A recent study by Jo \emph{et al.}~\cite{jo2025security} explored OAM as an encoding for 
high-dimensional MDI-QKD. Their primary contribution was a prepare–measure, 
three-party MDI protocol using qutrit OAM states and a linear-optical correlation 
measurement. For comparison, they also formulated an entanglement-based quantum secret 
sharing scheme using cyclic-entangled qutrit states, allowing a direct contrast between 
correlation-measurement and entanglement-distribution realizations while keeping the same 
OAM alphabet and detection model. The work highlighted a key constraint in 
high-dimensional optical systems: complete high-dimensional Bell-state or GHZ-state analysis 
is not achievable using only linear optics, motivating the use of cyclic-entangled 
projections instead. Their choice of OAM emphasized both the potential capacity and 
noise-resilience benefits of high-dimensional spatial encoding and the practical challenges 
of free-space OAM links, including turbulence and state-dependent diffraction.

While high-dimensional degrees of freedom such as OAM offer promising advantages for 
capacity scaling and robustness, a complete OAM-based DI-QKD system has not yet been 
demonstrated. Current research continues to explore individual components, such as
high-quality OAM entanglement sources, high-dimensional measurements,
and turbulence-resilient OAM transmission. However, an end-to-end
OAM-enabled DI-QKD implementation remains an open experimental goal.

\section{Conclusion}\label{conc}
OAM-encoded HD-QKD leverages the rich structure of spatial modes to increase information per detected photon and improve robustness to noise, offering capabilities that extend beyond conventional qubit-based protocols. Experimental progress has demonstrated the feasibility of OAM-based state preparation, free-space and fiber transmission, and high-dimensional detection, yet practical deployment still requires careful handling of mode-dependent diffraction, atmospheric turbulence, inter-modal coupling in fibers, detector imperfections, and hardware-induced loss.

System-level insights from existing demonstrations indicate that moderate alphabet sizes, low-loss mode converters, OAM-preserving channels, and carefully co-designed measurement architectures currently offer the best balance between complexity and performance. At the same time, recent advances, including fast AOs, machine-learning-based wavefront correction, multicore and air-core fibers, and compact integrated mode generators, are steadily relaxing long-standing limitations.

Looking forward, the integration of OAM encoding with emerging QKD paradigms such as twin-field architectures, spatial-mode multiplexed CV-QKD, and MDI/DI frameworks highlights a promising path toward long-distance, high-rate, and implementation-secure quantum communication. As experiments increasingly validate these ideas under realistic atmospheric and metropolitan conditions, OAM-encoded HD-QKD is poised to transition from laboratory demonstrations to practical and scalable quantum network technologies.


\section*{Acknowledgments}
This research was supported by Brain Pool program funded by the Ministry of Science and ICT through the National Research Foundation of Korea (RS-2025-25456394).

\section*{Conflict of interest}
The authors declare that there is no conflict of interest in this paper.



\bibliographystyle{elsarticle-num}

\vspace{-0.3cm}

\bibliography{sample2}

\end{document}